\documentclass[twocolumn]{aastex61}

\newcommand{\hei}{He~{\sc i}}

\newcommand{\ceiii}{Ce~{\sc iii}}
\newcommand{\ali}{Al~{\sc i}}
\newcommand{\fei}{Fe~{\sc i}}
\newcommand{\feii}{Fe~{\sc ii}}
\newcommand{\ki}{K~{\sc i}}
\newcommand{\cai}{Ca~{\sc i}}
\newcommand{\caii}{Ca~{\sc ii}}
\newcommand{\cri}{Cr~{\sc i}}
\newcommand{\crii}{Cr~{\sc ii}}
\newcommand{\srii}{Sr~{\sc ii}}
\newcommand{\mni}{Mn~{\sc i}}
\newcommand{\mnii}{Mn~{\sc ii}}

\newcommand{\ci}{C~{\sc i}}

\newcommand{\mgi}{Mg~{\sc i}}
\newcommand{\mgii}{Mg~{\sc ii}}
\newcommand{\tiii}{Ti~{\sc ii}}
\newcommand{\si}{S~{\sc i}}
\newcommand{\sii}{Si~{\sc i}}
\newcommand{\siii}{Si~{\sc ii}}

\newcommand{\nstars}{157}
\newcommand{\nstarsnew}{154}

\newcommand{\ntotalapstars}{238}
\newcommand{\bfield}{$\langle$\emph{B}$\rangle$}

\newcommand{\mlmf}{$\langle$\emph{B$_{\rm z}$}$\rangle$}

\shorttitle{APOGEE A\MakeLowercase{p}/B\MakeLowercase{p} Stars}
\shortauthors{Chojnowski et al.}

\begin{document}

\title{Discovery of Resolved Magnetically Split Lines in SDSS/APOGEE Spectra of {\nstars} A\MakeLowercase{p}/B\MakeLowercase{p} Stars}

\author{S. Drew Chojnowski}
\affiliation{Apache Point Observatory and New Mexico State University, P.O. Box 59, Sunspot, NM, 88349-0059, USA}

\author{Swetlana Hubrig}
\affiliation{Leibniz-Institut für Astrophysik Potsdam (AIP), An der Sternwarte 16, 14482 Potsdam, Germany}
\author{Sten Hasselquist}
\affiliation{Department of Physics \& Astronomy, University of Utah, 115 1400 E, Salt Lake City, UT 84112, USA}
\author{Fiorella Castelli}
\affiliation{Istituto Nazionale di Astrofisica, Osservatorio Astronomico di Trieste, via Tiepolo 11, 34143 Trieste, Italy}
\author{David G. Whelan}
\affiliation{Department of Physics, Austin College, 900 N. Grand Ave., Sherman, TX 75090, USA}
\author{Steven R. Majewski}
\affiliation{Department of Astronomy, University of Virginia, P.O. Box 400325, Charlottesville, VA 22904-4325, USA}
\author{Christian Nitschelm}
\affiliation{Centro de Astronom\'ia (CITEVA), Universidad de Antofagasta, Avenidos Angamos 601, Antofagasta 1270300, Chile}
\author{D.A. Garc\'ia-Hern\'andez}
\affiliation{Instituto de Astrof\'isica de Canarias (IAC), E-38205 La Laguna, Tenerife, Spain}
\affiliation{Universidad de La Laguna (ULL), Departamento de Astrof\'isica, E-38206 La Laguna, Tenerife, Spain}
\author{Keivan G. Stassun}
\affiliation{Department of Physics and Astronomy, Vanderbilt University, Nashville, TN 37235, USA}
\author{Olga Zamora}
\affiliation{Instituto de Astrof\'isica de Canarias (IAC), E-38205 La Laguna, Tenerife, Spain}
\affiliation{Universidad de La Laguna (ULL), Departamento de Astrof\'isica, E-38206 La Laguna, Tenerife, Spain}

\begin{abstract}
We report on magnetic field measurements of {\nstars} chemically peculiar A/B stars (Ap/Bp) based on resolved, magnetically split absorption lines present in $H$-band spectra provided by the Sloan Digital Sky Survey (SDSS)/Apache Point Observatory Galactic Evolution Experiment (APOGEE) survey. These stars represent the extreme magnetic end of a still-growing sample of $>$900 Ap/Bp stars selected among the APOGEE telluric standard stars as those with {\ceiii} absorption lines and/or literature Ap/Bp classifications. The lines most frequently resolved into their split components for these stars in the $H$-band pertain primarily pertain to {\ceiii}, {\crii}, {\fei}, {\mnii}, {\sii}, and {\caii}, in addition to one or more unidentified ions. Using mean magnetic field modulus ({\bfield}) estimates for transitions with known Land\'e factors, we estimate effective Land\'e factors for 5 {\ceiii} lines and 15 unknown lines and proceed to measure {\bfield} of {\nstars} stars, only 3 of which have previous literature estimates of {\bfield}. This 183\% increase in the number of Ap/Bp stars for which {\bfield} has been measured is a result of the large number of stars observed by SDSS/APOGEE, extension of high-resolution Ap/Bp star observations to fainter magnitudes, and the advantages of long wavelengths for resolving magnetically split lines. With {\bfield}$\sim$25 kG, the star 2MASS~J02563098+4534239 is currently the most magnetic star of the SDSS/APOGEE sample. Effective Land\'e factors, representative line profiles, and magnetic field moduli are presented. The validity of the results is supported using optical, high-resolution, follow-up spectra for 29 of the stars.
\end{abstract}


\section{Introduction} \label{intro}
Globally ordered magnetic fields are observed in roughly 10--20\% of intermediate and massive main sequence stars with spectral types between approximately F0 and B2. These stars, generally called chemically peculiar Ap and Bp stars (referred to as Ap stars in the rest of this Letter, for simplicity's sake), exhibit strong overabundances of certain iron peak elements and rare earths. The majority also exhibit underabundances of He, C, and O relative to solar abundances \citep[e.g.][]{ghazaryan18}, though the more massive Bp stars usually show overabundances of He and Si \citep[e.g.,][]{castro17}. These effects are caused by the disruption of normal diffusion processes by the strong magnetic fields. As stars of more than a few solar masses are expected to lack significant surface convection zones in which magnetic dynamos are generated in low-mass stars, the strong magnetic fields of Ap stars are unexpected and often attributed to fossil fields inherited from the collapsed gas clouds \citep{moss01}.

The magnetic fields of Ap stars are most frequently diagnosed through spectropolarimetric measurements of the mean longitudinal field, {\mlmf}, which is strongly dependent on the line of sight but can be obtained for the majority of Ap stars regardless of their rotation rates. However, an important group is formed by the Ap stars with low projected rotational velocities and strong kG magnetic fields. For these stars, conventional spectroscopy of sufficient resolution can be used to measure the mean magnetic field modulus, {\bfield}, based on the wavelength separations of resolved, magnetically split lines (RMSLs). Although somewhat dependent on the geometry of the observation, {\bfield} has the advantage of being primarily determined by the intrinsic stellar magnetic field strength. Prior to this work, 84 Ap stars with RMSLs were known \citep{mathys17}, and their study allowed the measurement of their magnetic fields and the establishment of a number of general properties. The star with the strongest magnetic field currently known, ``Babcock's Star'' (HD\,215441) is a B8V star with a surface dipole field strength of 34\,kG \citep{babcock60,preston69}.

The vast majority of past studies of Ap stars with RMSLs have relied on optical data of very slowly rotating stars (for which the Doppler effect does not serve to blend the split components), but spectroscopy at longer wavelengths provides several advantages in terms of resolving magnetically split lines. For example, the wavelength separation of magnetically split components is proportional to the square of the wavelength, such that for a fixed effective Land\'e factor, the spectral resolution required to resolve the split components of a line is significantly lower in the $H$-band than it would be in the optical. Further, the Doppler effect has a linear dependence on wavelength, such that magnetic splitting can be resolved in the $H$-band for stars that are rotating too fast for the splitting to be resolved in the optical.

In this Letter we present the analysis of a sub-sample of {\nstarsnew} new stars with RMSLs in the $H$-band, as well as of 3 stars for which RMSLs had been observed previously in the optical. The {\nstars} stars were identified among the largest-ever spectroscopic survey of Ap stars to date, being carried out by the Sloan Digital Sky Survey \citep[SDSS-III and SDSS-IV;][]{eisenstein11,blanton17} sub-survey known as the Apache Point Observatory Galactic Evolution Experiment \citep[APOGEE;][]{majewski17}. This near-tripling of the number of Ap stars with {\bfield} measurement is due partially to the volume and depth of the APOGEE survey. The {\nstars} RMSL stars were identified among a much larger sample of $\sim$1000 Ap/Am stars with multi-epoch APOGEE observations, and the 91/{\nstars} stars with $V$ magnitudes greater than 10 represents a 3000\% increase in the number of $V>10$ Ap stars for which {\bfield} has been measured.

\section{Data} \label{data}
\subsection{APOGEE $H$-band Spectroscopy}
This Letter focuses primarily on data from the two APOGEE instruments, which are 300-fiber, $R=22,500$, $H$-band spectrographs that operate on the Sloan 2.5m telescope \citep{gunn06} at Apache Point Observatory (APO) and on the Du Pont 2.5m telescope at Las Campanas Observatory (LCO). Throughout this Letter, we refer to vacuum wavelengths when discussing the $H$-band data. Each APOGEE instrument records most of the $H$-band (15145--16960\,{\AA}) on three detectors, with coverage gaps between 15800--15860\,{\AA} and 16430--16480\,{\AA} and with each fiber having a $\sim$2\arcsec diameter on-sky field of view. Total exposure times in each observation are about one hour, and the the 2$^{\circ}$--3$^{\circ}$ diameter APOGEE fields are usually observed multiple times on different nights, months, or years, to accumulate signal for fainter targets and to check for radial velocity variability. Individual spectra are ultimately combined into high signal-to-noise ratio (S/N) spectra for chemical abundance analysis. A detailed description of the APOGEE survey was presented by \citet{majewski17}, and the data reduction process has been described in \citet{nidever15}. 

As discussed in \citet{zasowski13,zasowski17}, 35/300 APOGEE fibers are placed on blank sky positions to facilitate the removal of airglow emission features from science spectra, and a further 15--35/300 fibers are placed on quasi-randomly selected hot stars that facilitate removal of telluric absorption features from science spectra. The latter are generally the bluest and brightest available stars in the field and are restricted to a magnitude range of roughly $6.5<H<11.0$. They are selected mostly based on raw Two Micron All-Sky Survey \citep[2MASS;][]{skrutskie06} $J-K$ color but also with spatial restrictions to account for variation of the telluric absorption across the field. To date, the APOGEE instruments have observed $>$40,000 telluric standard stars, permitting the serendipitous discovery of exotic objects including highly magnetized OB stars \citep{eikenberry14}, Be stars \citep{choj15}, as well as a large sample of $>$900 Ap stars, some of which are the focus of this Letter.

\begin{figure*}[ht!]
\epsscale{1.0}
\plotone{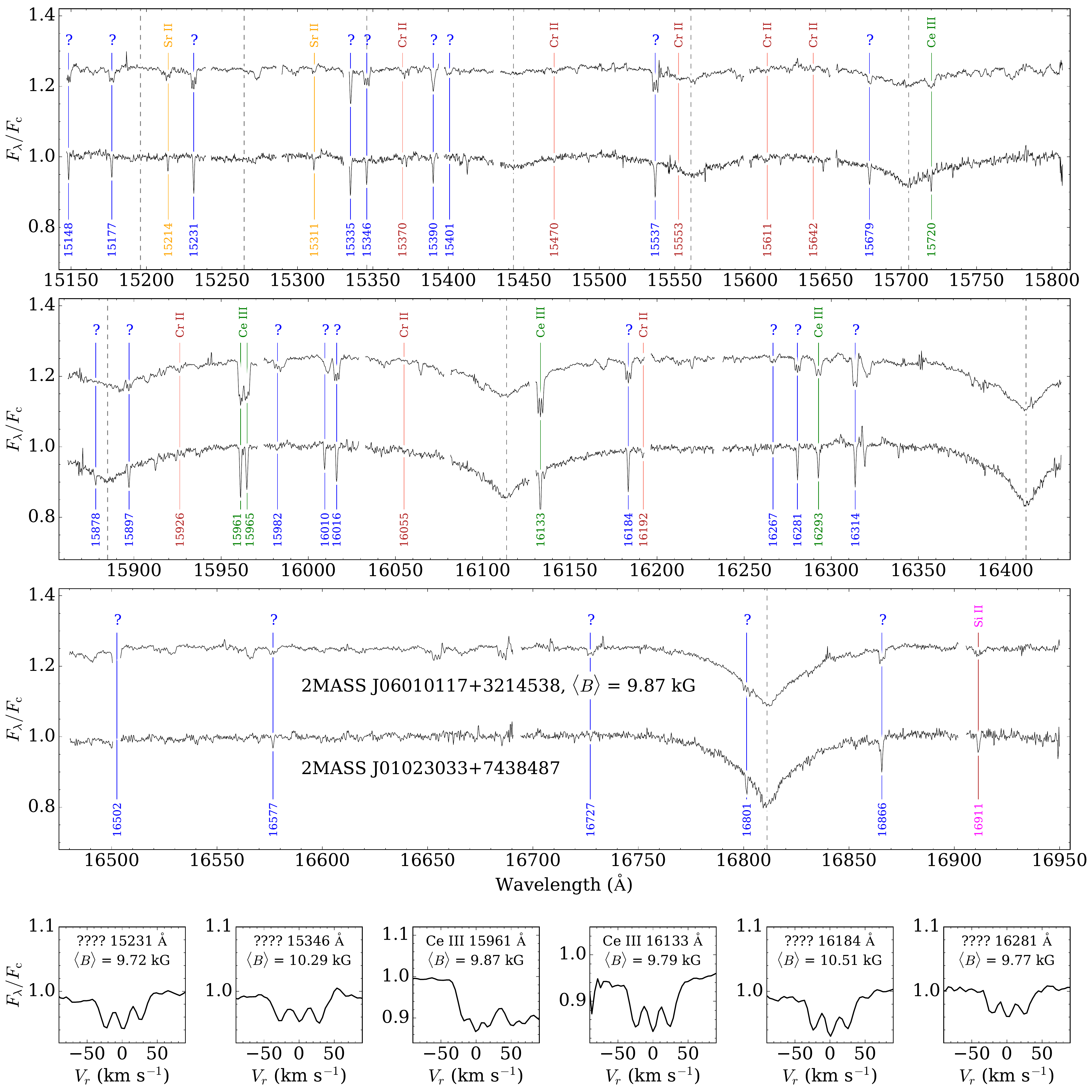}
\caption{\textbf{Top three rows:} APOGEE spectra of 2MASS\,J06010117+3214538 and 2MASS\,J01023033+7438487, two stars exhibiting the unknown lines discussed in Section~\ref{lineids}, with the lines being magnetically split for the former. Blue vertical lines indicate the unknown lines (for which the labels are the vacuum rest wavelengths in {\AA}) and green lines indicate {\ceiii}. A few other species in common between the two stars are labeled. Strong airglow lines have been masked from the spectra, including one that typically distorts the profile of the unknown line at 16502~{\AA}. \textbf{Bottom row:} a closer view of magnetically split lines in the spectrum of 2MASS\,J06010117+3214538. The feature redward of {\ceiii}~15961\,{\AA} is a blend of {\sii}~15964\,{\AA} and {\ceiii}~15965\,{\AA}. The mean magnetic field modulus obtained from the measured splitting of each line is given. \label{example_2M06010117}}
\end{figure*}

\subsection{ARCES Optical Spectroscopy}
We also obtained optical spectra of 29 stars using the Astrophysical Research Consortium Echelle Spectrograph \citep[ARCES;][]{wang03} on the ARC 3.5m telescope at APO. ARCES covers the full optical spectrum (3500--10000 {\AA}) at a resolution of $R$=31,500. We used standard Image Reduction and Analysis Facility (IRAF\footnote{IRAF is distributed by the National Optical Astronomy Observatories, which are operated by the Association of Universities for Research in Astronomy, Inc., under cooperative agreement with the National Science Foundation.}) echelle data reduction techniques, including 2D to 1D extraction, bias subtraction, scattered light and cosmic ray removal, flat-field correction, wavelength calibration via Thorium--Argon lamp exposures, as well as continuum normalization and merging of orders.

\section{Sample Selection} \label{sample}
The APOGEE Ap star sample currently consists of 986 stars that APOGEE has observed together more than 5500 times, with the about 85\% of the stars having been identified among the APOGEE telluric standard stars via an algorithm designed to search for the presence of the doubly ionized Cerium ({\ceiii}) lines that are usually the strongest absorption features aside from the broad hydrogen Bracket series lines. The other 15\% of the stars lack the {\ceiii} lines and were instead added to the sample based on literature Ap/Bp/Am classifications \citep[for detailed definitions of these spectral classes, see Chapter 5.2 of][]{graycorbally}. These 986 Ap/Bp/Am stars therefore account for just over 2\% of the APOGEE telluric standard stars, but this should be taken as a lower limit to the true fraction. 

Whereas the overall sample will be discussed in future work, this Letter deals with the {\nstars} stars (886 total spectra) for which RMSLs were visually identified in one or more APOGEE spectrum. For the majority of the {\nstars} stars, RMSLs appear in multiple observations such that absorption lines in the combined spectra clearly exhibit splitting. However, 17/{\nstars} stars have only been observed once to date, and in another 49/{\nstars} cases, the RMSLs were noticed and measured in an individual spectrum rather than the combined spectrum. Usually this was due to one of the individual spectra having significantly higher S/N than the others, but in some cases, it was almost certainly due to temporal variability of the magnetic field strength. 

\begin{deluxetable*}{lcrccrccrcc}
\tablecaption{Effective Land\'e $g$ Factors. \label{landegtable}}
\tablehead{
\colhead{Ion} & \colhead{$\lambda_{\rm vac}$} & \colhead{log($gf$)} & \colhead{$J_{\rm low}$} & \colhead{$g_{\rm low}$} & \colhead{$E_{\rm low}$} & \colhead{$J_{\rm high}$} & \colhead{$g_{\rm high}$} & \colhead{$E_{\rm high}$} & \colhead{$g_{\rm eff}$} & \colhead{N$_{\rm stars}$} \\
\colhead{}    & \colhead{({\AA})}             & \colhead{}          & \colhead{}              & \colhead{}              & \colhead{(eV)}          & \colhead{}               & \colhead{}               & \colhead{(eV)}           & \colhead{}              & \colhead{}
}
\startdata
{\ci} & 16009.273 & -0.090 & 2.0 & 1.000 & 9.631 & 3.0 & 1.001 & 10.406 & $1.002\pm0.050$ & 3 \\
{\ci} & 16026.078 & -0.140 & 2.0 & 1.000 & 9.631 & 3.0 & 1.074 & 10.405 & $1.148\pm0.050$ & 6 \\
{\ci} & 16895.031 & 0.568 & 2.0 & 1.000 & 9.003 & 3.0 & 0.978 & 9.736 & $0.956\pm0.050$ & 19 \\
{\mgi} & 15753.291 & -0.060 & 1.0 & 1.501 & 5.932 & 2.0 & 1.167 & 6.719 & $1.000\pm0.050$ & 2 \\
{\mgi} & 15770.150 & 0.380 & 2.0 & 1.501 & 5.933 & 3.0 & 1.334 & 6.719 & $1.167\pm0.050$ & 8 \\
{\mgii} & 16764.796 & 0.480 & 0.5 & 0.666 & 12.083 & 1.5 & 0.800 & 12.822 & $0.834\pm0.050$ & 9 \\
{\mgii} & 16804.520 & 0.730 & 1.5 & 1.334 & 12.085 & 2.5 & 1.200 & 12.822 & $1.099\pm0.050$ & 8 \\
{\ali} & 16723.524 & 0.152 & 0.5 & 0.666 & 4.085 & 1.5 & 0.800 & 4.827 & $0.834\pm0.050$ & 6 \\
{\ali} & 16755.140 & 0.408 & 1.5 & 1.334 & 4.087 & 2.5 & 1.200 & 4.827 & $1.099\pm0.050$ & 10 \\
{\sii} & 15888.794 & -0.740 & 1.0 & 0.508 & 5.954 & 1.0 & 1.472 & 6.734 & $0.990\pm0.050$ & 2 \\
\enddata
\tablecomments{Only the first 10 rows are shown. The full version of this table is available in machine-readable format in the online version.}
\end{deluxetable*}

\section{Line Identification} \label{lineids}
We used the APOGEE linelist \citep{shetrone15} to identify the majority of absorption lines present in the $H$-band spectra. Atomic data for the {\ceiii} lines was obtained from a {\ceiii} linelist computed by \citet{biemont02}. This linelist was previously used by \citet[][provided to them as private communication by Biemont et al.]{hubrig12}, who were the first to confirm the presence of {\ceiii} lines in $H$-band spectra of Ap stars. For the optical follow-up spectra, we relied on the Kurucz linelist\footnote{1995 Atomic Line Data (R.L. Kurucz and B. Bell) Kurucz CD-ROM No. 23. Cambridge, Mass.: Smithsonian Astrophysical Observatory.} for line identification. 

One or more ions are clearly missing from our $H$-band linelist, as indicated by several unidentified lines exhibiting RMSLs in the spectra of numerous stars. These features are usually detected simultaneously and they are occasionally the only metallic absorption features beyond {\ceiii}. They are probably due to some heavy element that cannot be identified owing to the limited availability of atomic data for $\lambda>10000$~{\AA}. The strongest of these unknown lines is the 16184~{\AA} line, for which magnetically split components are observed in $\sim$31\% of the larger Ap star sample. We used a sub-sample of 20 narrow-lined Ap stars exhibiting the unknown lines to measure accurate wavelengths for the features. Figure~\ref{example_2M06010117} displays the full APOGEE spectrum of one such narrow-lined star (2MASSJ\,01023033+7438487) along with a star for which the unknown lines are magnetically split (2MASSJ\,06010117+3214538).

\begin{figure}
\epsscale{1.0}
\plotone{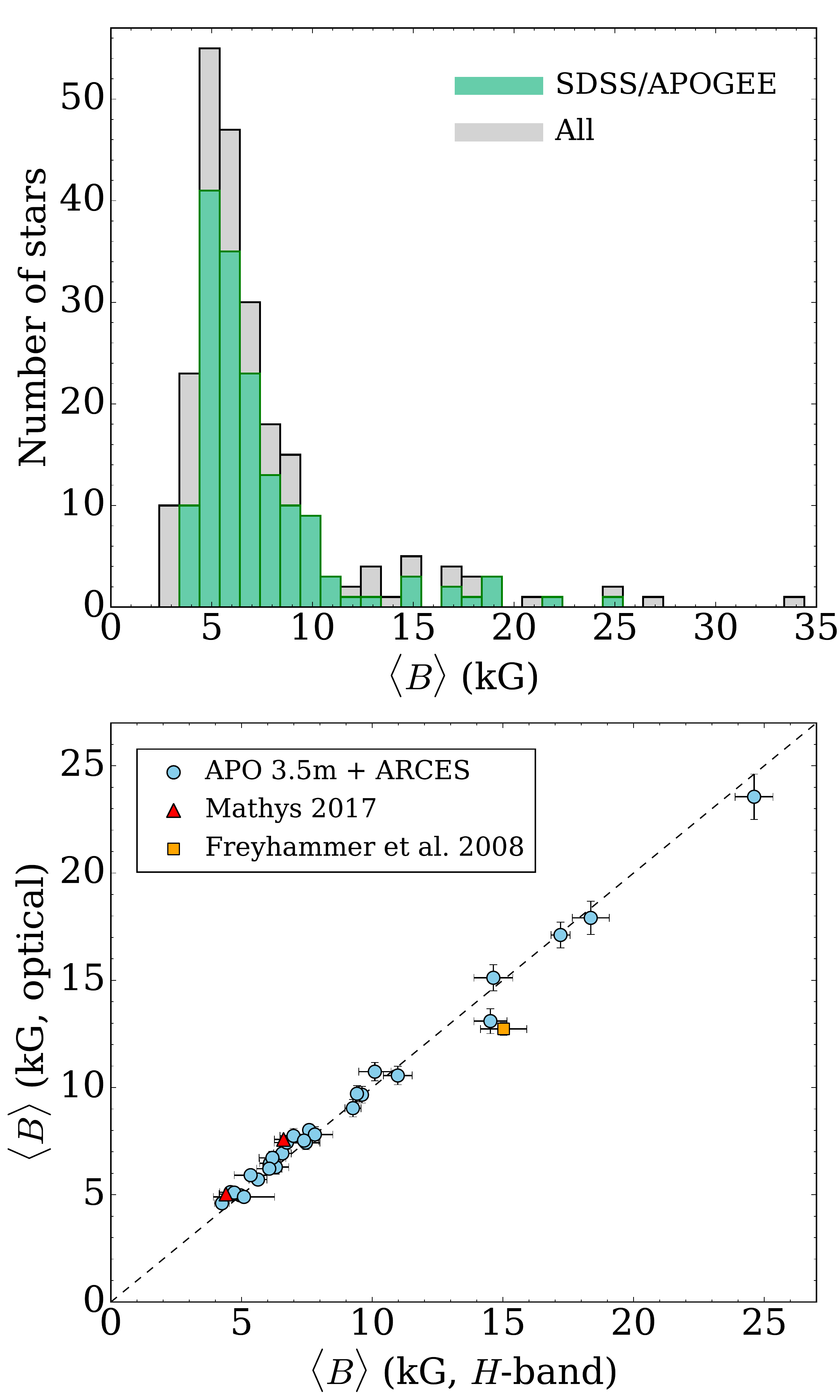}
\caption{\textbf{Upper panel:} histogram of the magnetic field modulus for all {\ntotalapstars} Ap stars with resolved, magnetically split lines. The new additions from SDSS/APOGEE are shaded in green. \textbf{Lower panel:} comparison of {\bfield} as measured in the $H$-band versus the optical. The dashed line indicates perfect agreement. \label{histB}}
\end{figure}

The {\ceiii} lines (15961, 15965, 16133~{\AA}, and in cases of strong {\ceiii}, also 16292 and 15720~{\AA}) are present in 100\% of the sample at hand, with the 16133~{\AA} line exhibiting RMSLs for 63\% of the sample and with the weaker 15961~{\AA} and 16292~{\AA} lines being split for 43\% and 32\% of the sample, respectively. The {\ceiii}~15965~{\AA} line is often blended with the strongest {\sii} line covered (15964~{\AA}), such that the splitting of {\ceiii}~15965~{\AA} can only be measured when the presence of {\sii} lines can be ruled out by confirming absence of {\sii}~15893~{\AA} and 16685~{\AA}. The fraction of stars with {\ceiii}~16133~{\AA} measurements would be even higher were it not for a strong airglow feature (16129~{\AA}) that is often poorly subtracted and may coincide with the blue wing of the {\ceiii} line depending on the stellar radial velocity.

{\crii} is the next most frequently magnetically split ion, with the splitting of the 15370 and 15470~{\AA} lines being measurable in 50\% and 39\% of the sample, respectively. Other particularly useful lines include {\fei}~15299~{\AA} (34\%), {\fei}~15626~{\AA} (31\%), {\fei}~16491~{\AA} (27\%), {\sii}~16685~{\AA} (25\%), as well as the unidentified line at 16184~{\AA} (31\%). Unlike the situation in the optical, where Ap stars typically exhibit numerous {\feii} and {\tiii} lines, no {\feii} lines are confidently detected in the spectra of any stars, and {\tiii} detections are quite rare. The star 2MASS\,J23102121+4717017 is the only available example where more than two magnetically split {\tiii} lines were resolved.

\section{Mean Magnetic Field Modulus Measurement} \label{measure}
Given measurements of the wavelength separations of magnetically split components in a given absorption line and knowledge of the Land\'e $g$ factors of the associated energy levels, it is straightforward to compute the mean magnetic field modulus {\bfield} in units of Gauss according to 

\begin{equation} \label{eq1}
\langle B \rangle = \frac{\Delta\lambda}{{k}~{g_{\rm eff}}~{\lambda_{0}^{2}}}
\end{equation}

\noindent Here, $\Delta\lambda$ is the separation in {\AA} between an outer and the central split component or else half of the separation between outer components in a triplet profile. $\lambda_{0}$ is the rest wavelength of the line in {\AA} and $k$ is a constant, $4.67\times10^{-13}$~{\AA}$^{-1}$~G$^{-1}$. It is important to note that although we have only analyzed spectral lines exhibiting apparent triplet profiles like those shown in the lower panels of Figure~\ref{example_2M06010117}, the Zeeman patterns \citep[see][]{mathys90} of the $H$-band lines have not been investigated and Equation~\ref{eq1} is therefore an approximation.

The effective Land\'e factor ($g_{\rm eff}$) of a given transition is calculated using the total angular momentum quantum numbers ($J$) and Land\'e factors ($g$) of the lower (1) and upper level (2) as follows:

\begin{equation} \label{eq2}
g_{\rm eff} = \frac{1}{2} (g_{\rm 1}+g_{\rm 2}) + \frac{1}{4} (g_{\rm 1}-g_{\rm 2})\Big[J_{\rm 1}(J_{\rm 1}+1)-J_{\rm 2}(J_{\rm 2}+1)\Big],
\end{equation}

\noindent where in the case of $J_{1}=J_{2}$, $g_{\rm eff}$ is simply the average of the Land\'e factors of the lower and upper levels.

To measure $\Delta\lambda$, we either fit Gaussians to the absorption features or simply estimated their positions visually. This was done interactively using the $splot$ program in IRAF. For each star, the splitting was measured for as many lines as possible and the results were averaged, with assumed or derived errors on $g_{\rm eff}$ being propagated appropriately. On average, measurements from $\sim$13 lines per star were used, but the range of number of lines used is quite large. In a few cases, the magnetic splitting could only be measured in two to three lines, often because the only metallic lines present are {\ceiii}. In the case of 2MASS\,J06365367+0118455 however, the splitting of 61 lines was measured.


The $J$ and $g$ values used here were taken from the Kurucz linelist, and we assumed a constant uncertainty of 0.02 for the $g_{\rm eff}$ of lines with existing laboratory data. However, laboratory data are unavailable for the {\ceiii} lines and for the unknown lines (see Section~\ref{lineids}), such that it was necessary to estimate the $g_{\rm eff}$ of these lines. To do so, we defined a sub-sample of 34/{\nstars} stars with high-S/N spectra, particularly well-resolved magnetically split features, and measurements of lines with known $g_{\rm eff}$ in addition to {\ceiii} and the unknown lines. The lines with known $g_{\rm eff}$ were used to calculate preliminary {\bfield} measurements for the 34 stars. These values were then used derive average estimates of $g_{\rm eff}$ for the {\ceiii} and unknown lines. 

Table~\ref{landegtable} summarizes the 149 lines used for measurement of {\bfield}, providing the vacuum wavelength, log($gf$), energy levels, and $g_{\rm eff}$ for each line, as well as the number of stars for which magnetic splitting was measured. Overall, we find relatively low $g_{\rm eff}$ for {\ceiii} and the unknown lines, which together average $g_{\rm eff}\sim$1.07. The weak {\ceiii}~15720\,{\AA} line has the highest estimated value at $g_{\rm eff}\sim$1.39, but unfortunately this line is rarely observed. 

\subsection{$H$-band Results}
The upper panel of Figure~\ref{histB} shows the distribution of measured {\bfield} of all Ap stars known to exhibit RMSLs, with the new discoveries presented in this Letter shaded in green and with those from \citet{mathys17} shaded in gray. The average for the APOGEE sample is {\bfield}$\sim$7.1\,kG, while the average for all stars including those of \citet{mathys17} is a bit higher at {\bfield}$\sim$7.3\,kG due to the inclusion of Babcock's Star. The distribution of {\bfield} measured using APOGEE spectra ranges from $\sim$3.6 kG in the case of 2MASS\,J02463240+0001145 up to $\sim$25 kG in the case of 2MASS\,J02563098+4534239. A total of 19 new examples of stars with {\bfield}$>$10\,kG are identified, thus filling in the previous gap between 10--11\,kG. 

The lower limit on {\bfield} of $\sim$3.5\,kG measurable from the APOGEE spectra is a consequence of resolving power, whereby the smallest splitting that we can measure is about 0.6~{\AA}. For a star with strong {\mni}~15222~{\AA} ($g_{\rm eff}$=1.969), splitting should be visible down to {\bfield} of just below 3\,kG. No such examples have been found, however, meaning that it is not possible to investigate the notion of an intrinsic lower limit to measurable field strength of {\bfield}$\sim2$\,kG \citep{mathys17}.

Figure~\ref{ZsplitMontage1} shows an example of a magnetically split line for each of the {\nstars} stars, generally showing the most clearly resolved line. In the cases where {\ceiii}~16133\,{\AA} is displayed, narrow spikes blueward of the line are residuals from the imperfect removal of the aforementioned strong airglow line at 16129\,{\AA}.

\begin{deluxetable*}{clrrcrlrrrr}
\tablecaption{Magnetic Field Modulus Estimates. \label{Bfieldtable}}
\tabletypesize{\scriptsize}
\tablehead{
\colhead{2MASS ID} & \colhead{Other ID} & \colhead{$V$}   & \colhead{$H$}   & \colhead{$N_{\rm spectra}$} & \colhead{S/N} & \colhead{Spec. Type} & \colhead{{\bfield} (kG)} & \colhead{$N_{\rm lines}$} & \colhead{{\bfield} (kG)} & \colhead{$N_{\rm lines}$} \\
\colhead{}         & \colhead{}         & \colhead{(mag)} & \colhead{(mag)} & \colhead{}                  & \colhead{}    & \colhead{}           & \colhead{$H$-band}       & \colhead{$H$-band}        & \colhead{Optical}        & \colhead{Optical}        
}
\startdata
00033808+7018217 & HD 225114 & 8.10 & 8.19 & 12 & 726 & A0p SrCrSi\tablenotemark{4} & $7.46\pm0.54$ & 8 & $7.43\pm0.32$\tablenotemark{4} & 36 \\
00102704+7337035 & TYC 4306-1062-1 & 10.84 & 10.07 & 9 & 537 & \nodata & $4.17\pm0.28$ & 16 & \nodata & \nodata \\
00283062+6947472 & TYC 4299-696-1 & 10.53 & 9.45 & 12 & 682 & \nodata & $7.92\pm0.89$ & 4 & \nodata & \nodata \\
00284036+8418313 & TYC 4615-2915-1 & 10.96 & 10.54 & 14 & 332 & A2V\tablenotemark{2} & $5.10\pm0.22$ & 17 & \nodata & \nodata \\
00323366+5512530 & HD 2887 & 9.79 & 8.17 & 13 & 1793 & A2:p SrCr\tablenotemark{4} & $4.95\pm0.24$ & 25 & $4.97\pm0.22$\tablenotemark{4} & 27 \\
00331847+5716167 & TYC 3662-378-1 & 10.70 & 10.37 & 6 & 481 & \nodata & ($8.53\pm0.79$) & 5 & \nodata & \nodata \\
00564145+5739255 & TYC 3676-505-1 & 11.19 & 10.96 & 3 & 86 & \nodata & $18.49\pm0.76$ & 6 & \nodata & \nodata \\
00584870+6240562 & TYC 4021-632-1 & 11.22 & 10.45 & 6 & 411 & A2:p Eu\tablenotemark{4} & $4.25\pm0.29$ & 15 & $4.60\pm0.19$\tablenotemark{4} & 22 \\
01052242+5010296 & TYC 3271-1597-1 & 10.49 & 10.22 & 3 & 266 & \nodata & ($6.79\pm1.13$) & 3 & \nodata & \nodata \\
01184266+5844433 & TYC 3681-1528-1 & 10.49 & 9.99 & 4 & 355 & B8\tablenotemark{2} & $9.67\pm0.50$ & 9 & \nodata & \nodata \\
\enddata
\tablecomments{The full version of this table is available in machine-readable format in the online version.}
\tablenotetext{1}{\citet{renson09}}
\tablenotetext{2}{\citet{skiff14}}
\tablenotetext{3}{SIMBAD}
\tablenotetext{4}{Own data (ARC 3.5m/ARCES)}
\tablenotetext{5}{\citet{mathys17}}
\tablenotetext{6}{\citet{freyhammer08}}
\end{deluxetable*}

\subsection{Optical Results}
The same procedure described in Section~\ref{measure} was applied to the 29 optical follow-up spectra obtained with the ARC 3.5m telescope and ARCES spectrograph in order to provide a sanity check on the $H$-band measurements of {\bfield}. As expected, the stars exhibit numerous RMSLs in the optical. For the 19/29 stars with {\bfield}$<$8\,kG, we relied primarily on limited numbers of long wavelength lines (7000--10200~{\AA}), while for the 10/29 stars with {\bfield}$>$9\,kG, we measured hundreds of RMSLs for each star, keeping the 50 lines nearest to the average. Another 3 of the {\nstars} stars  were included in the \citet{mathys17} sample of stars for which magnetically split lines had been previously resolved in optical spectra. 

As demonstrated in the lower panel of Figure~\ref{histB}, we find good agreement overall between the $H$-band and optical measurements, with the results agreeing to within a kG in most cases. The most discrepant results pertain to the star 2MASSJ\,07123042-2103537 (HD\,55540), where we find {\bfield}=15.0\,kG despite \citet{freyhammer08} having found {\bfield}=12.7\,kG. The three available APOGEE spectra of this star allow us to confirm that not only is the difference due to temporal variability of the observed {\bfield}, but also that the star is indeed radial velocity variable as pointed out by \citet{freyhammer08}. 

We also used the optical spectra to estimate traditional spectral types based on comparison of the target stars to a sample of 30 bright A and B stars with spectra in the ELODIE Archive \citep{prugniel01}. Stars with clear detections of {\hei}~4471\,{\AA} and weak or missing {\fei}, {\mgi}, {\cai}, and {\mni} lines were assigned temperature classes of B8 or B9, while the A stars were classified based on the equivalent widths of temperature-dependent lines ($e.g.,$ {\fei}~4045\,{\AA}, {\mgi}~5183\,{\AA}, {\cai}~4227\,{\AA}, {\mni}~4031\,{\AA}) as well the ratios of neutral versus singly ionized lines ($e.g.,$ {\mgi}/{\mgii} and {\fei}/{\feii}). Due in part to difficulty finding the continuum around the wings of the hydrogen Balmer series lines in the ARCES camera's small wavelength ranges per order, luminosity classes were not estimated. Anomalously strong absorption lines from Si, Cr, Sr, and Eu are specifically flagged.

Table~\ref{Bfieldtable} summarizes the $H$-band and optical magnetic field modulus measurements, giving 2MASS designations, alternate identifiers, $V$ magnitudes from UCAC4 \citep{zacharias13}, $H$ magnitudes from 2MASS, the number of APOGEE spectra of each star, the S/N ratio of the combined spectrum, a literature or estimated spectral type if available, the magnetic field modulus {\bfield}, and the number of spectral lines used to measure {\bfield}. In the cases where optical measurements were available, the final columns of Table~\ref{Bfieldtable} provide the associated {\bfield} measurement and the number of spectral lines used. Less reliable measurements of {\bfield}, usually due to large scatter between different lines and/or measurements near the resolution limit of APOGEE are given in parentheses in Table~\ref{Bfieldtable}.

\section{Conclusions}
Although spectroscopic observations at long wavelengths such as in the $H$-band help to maximize the $\Delta\lambda$ term of Equation~\ref{eq1} and to lessen the impact of blending due to the Doppler effect, this Letter represents the first serious exploitation of this wavelength regime for investigating the magnetic fields of Ap stars. The APOGEE sample of Ap stars with RMSLs represents a 183\% increase in the number of Ap stars with {\bfield} measurements, bringing the total to {\ntotalapstars} stars. As the SDSS/APOGEE survey is still ongoing, this number is expected to increase. Detailed characterization of the now much larger sample of Ap stars with RMSLs may help to shed light on the origin of magnetic fields in stars with radiative envelopes as well as the relations between magnetism and other factors such as binarity and rotation. In particular, an investigation of the stellar parameters and chemical abundances of the sample would be a worthwhile effort, as would be high-resolution spectroscopic monitoring to measure the average value of {\bfield} and the rotational period.

As demonstrated in the histogram of magnetic field modulus presented in Figure~\ref{histB}, more than 50\% of Ap stars with magnetically split lines possess magnetic fields in the range of 4--7\,kG. Despite the identification among the APOGEE sample of 19 new examples of stars with {\bfield}$>$10\,kG, led by 2MASS\,J02563098+4534239 ({\bfield}=25\,kG), the high-{\bfield} tail of the distribution remains capped by the 34\,kG magnetic field of Babcock's Star \citep[HD\,215441;][]{babcock60}. The only other non-degenerate star known to approach 30\,kG is HD\,75049 \citep{elkin10}, for which {\bfield} ranges from 24 to 30\,kG over an orbital period. Among the more massive magnetic stars, which are He-rich early-type B stars, the strongest known magnetic field is only 21\,kG \citep{hubrig17}. We can speculate that the magnetic field value of 34\,kG likely represents a critical field strength above which stable magnetic fields do not exist in Ap stars. Such a limit would be probably be related to the restricted range of seed fields, of the order of milli-Gauss, that are observed in star-forming regions \citep{han07}.

\vspace{0.5cm}
\scriptsize{\emph{Acknowledgements.} Funding for the Sloan Digital Sky Survey IV has been provided by the Alfred P. Sloan Foundation, the U.S. Department of Energy Office of Science, and the Participating Institutions. SDSS acknowledges support and resources from the Center for High-Performance Computing at the University of Utah. The SDSS web site is www.sdss.org.

SDSS is managed by the Astrophysical Research Consortium for the Participating Institutions of the SDSS Collaboration including the Brazilian Participation Group, the Carnegie Institution for Science, Carnegie Mellon University, the Chilean Participation Group, the French Participation Group, Harvard-Smithsonian Center for Astrophysics, Instituto de Astrofísica de Canarias, The Johns Hopkins University, Kavli Institute for the Physics and Mathematics of the Universe (IPMU) / University of Tokyo, Lawrence Berkeley National Laboratory, Leibniz Institut für Astrophysik Potsdam (AIP), Max-Planck-Institut für Astronomie (MPIA Heidelberg), Max-Planck-Institut für Astrophysik (MPA Garching), Max-Planck-Institut für Extraterrestrische Physik (MPE), National Astronomical Observatories of China, New Mexico State University, New York University, University of Notre Dame, Observatório Nacional / MCTI, The Ohio State University, Pennsylvania State University, Shanghai Astronomical Observatory, United Kingdom Participation Group, Universidad Nacional Autónoma de México, University of Arizona, University of Colorado Boulder, University of Oxford, University of Portsmouth, University of Utah, University of Virginia, University of Washington, University of Wisconsin, Vanderbilt University, and Yale University.

D.A.G.H. and O Z. acknowledge support from the State Research Agency (AEI) of the Spanish Ministry of Science, Innovation and Universities (MCIU) and the European Regional Development Fund (FEDER) under grant AYA2017-88254-P.

We thank the referee for numerous useful suggestions that improved this Letter.

\begin{figure*}
\epsscale{1.0}
\plotone{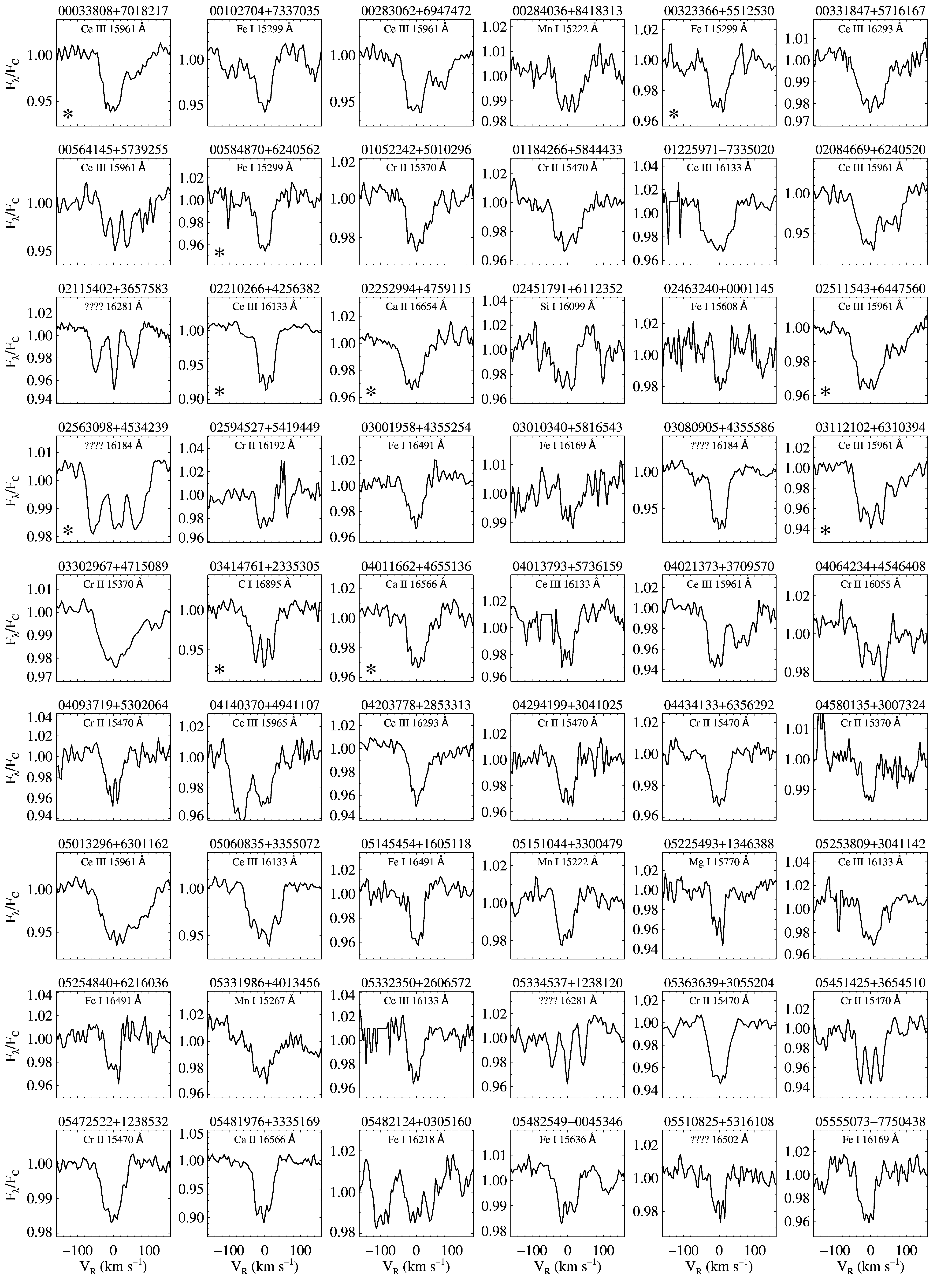}
\caption{Montage of Zeeman split lines in the SDSS/APOGEE spectra for all {\nstars} stars. 2MASS designations are provided above each panel, and asterisks in the lower left corners of some panels indicate that the presence of resolved, magnetically split lines was confirmed via higher-resolution optical spectra. \label{ZsplitMontage1}}
\end{figure*}

\begin{figure*}
\epsscale{1.0}
\plotone{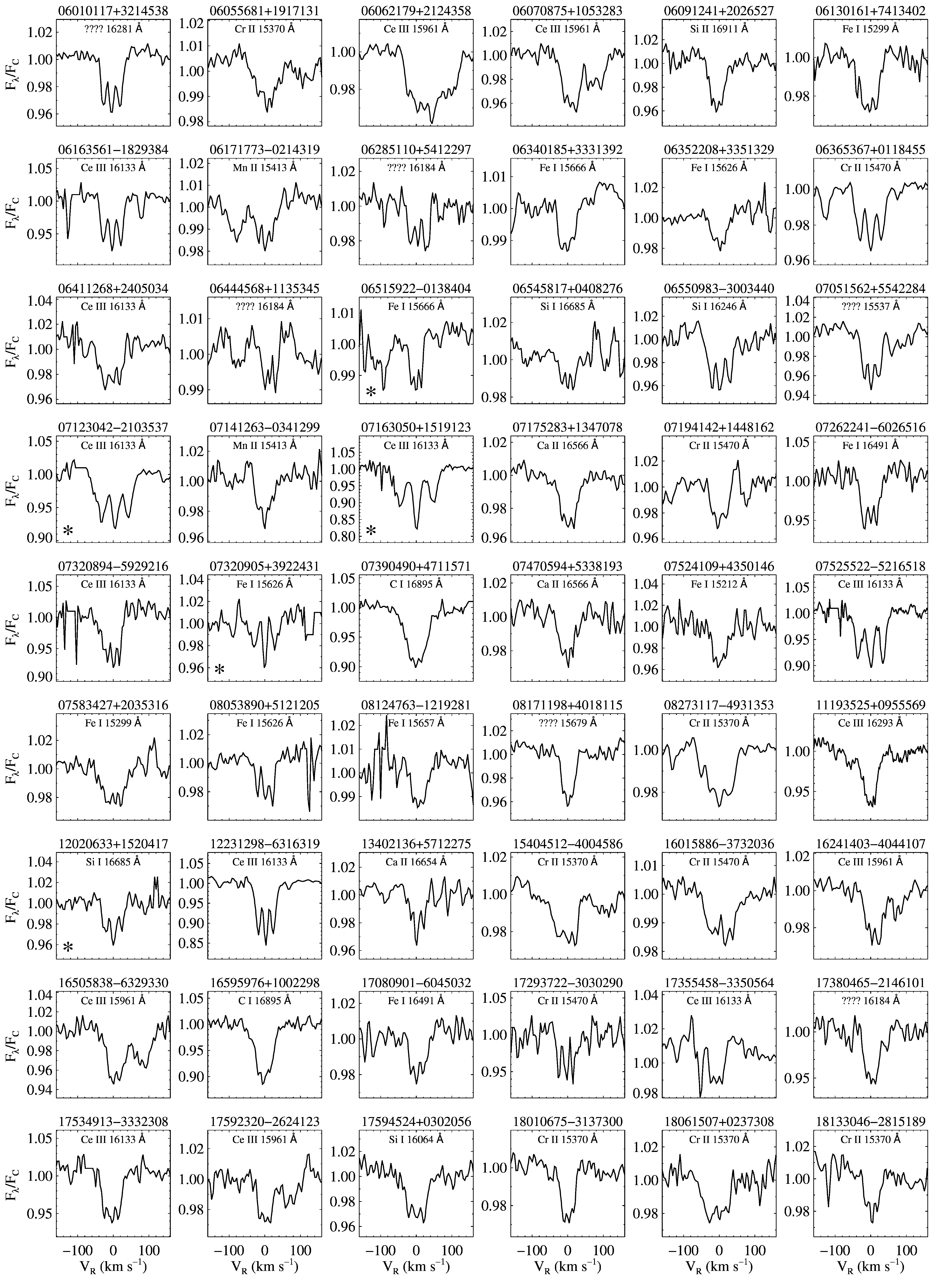}
\caption{Figure~\ref{ZsplitMontage1} continued.\label{ZsplitMontage2}}
\end{figure*}

\begin{figure*}
\epsscale{1.0}
\plotone{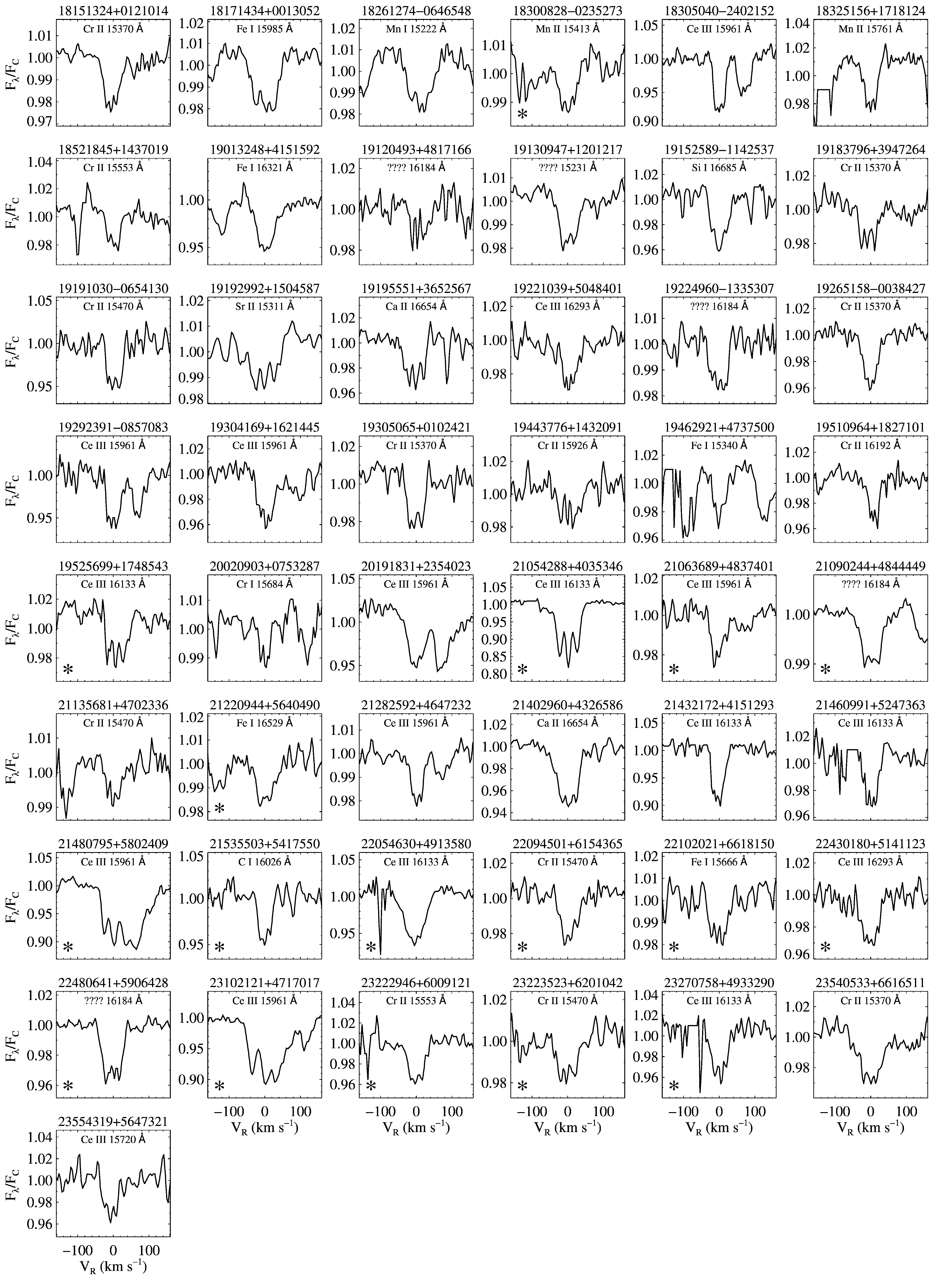}
\caption{Figure~\ref{ZsplitMontage1} continued.\label{ZsplitMontage3}}
\end{figure*}

\clearpage

\begin{longrotatetable}
\begin{deluxetable*}{lcrccrccrcc}
\tablecaption{Effective Land\'e $g$ Factors.}
\tablehead{
\colhead{Ion} & \colhead{$\lambda_{\rm vac}$} & \colhead{log($gf$)} & \colhead{$J_{\rm low}$} & \colhead{$g_{\rm low}$} & \colhead{$E_{\rm low}$} & \colhead{$J_{\rm high}$} & \colhead{$g_{\rm high}$} & \colhead{$E_{\rm high}$} & \colhead{$g_{\rm eff}$} & \colhead{N$_{\rm stars}$} \\
\colhead{}    & \colhead{({\AA})}             & \colhead{}          & \colhead{}              & \colhead{}              & \colhead{(eV)}          & \colhead{}               & \colhead{}               & \colhead{(eV)}           & \colhead{}              & \colhead{}
}
\startdata
{\ci} & 16009.273 & -0.090 & 2.0 & 1.000 & 9.631 & 3.0 & 1.001 & 10.406 & $1.002\pm0.050$ & 3 \\
{\ci} & 16026.078 & -0.140 & 2.0 & 1.000 & 9.631 & 3.0 & 1.074 & 10.405 & $1.148\pm0.050$ & 6 \\
{\ci} & 16895.031 & 0.568 & 2.0 & 1.000 & 9.003 & 3.0 & 0.978 & 9.736 & $0.956\pm0.050$ & 19 \\
{\mgi} & 15753.291 & -0.060 & 1.0 & 1.501 & 5.932 & 2.0 & 1.167 & 6.719 & $1.000\pm0.050$ & 2 \\
{\mgi} & 15770.150 & 0.380 & 2.0 & 1.501 & 5.933 & 3.0 & 1.334 & 6.719 & $1.167\pm0.050$ & 8 \\
{\mgii} & 16764.796 & 0.480 & 0.5 & 0.666 & 12.083 & 1.5 & 0.800 & 12.822 & $0.834\pm0.050$ & 9 \\
{\mgii} & 16804.520 & 0.730 & 1.5 & 1.334 & 12.085 & 2.5 & 1.200 & 12.822 & $1.099\pm0.050$ & 8 \\
{\ali} & 16723.524 & 0.152 & 0.5 & 0.666 & 4.085 & 1.5 & 0.800 & 4.827 & $0.834\pm0.050$ & 6 \\
{\ali} & 16755.140 & 0.408 & 1.5 & 1.334 & 4.087 & 2.5 & 1.200 & 4.827 & $1.099\pm0.050$ & 10 \\
{\sii} & 15888.794 & -0.740 & 1.0 & 0.508 & 5.954 & 1.0 & 1.472 & 6.734 & $0.990\pm0.050$ & 2 \\
{\sii} & 15892.751 & -0.030 & 1.0 & 1.005 & 5.082 & 1.0 & 0.996 & 5.862 & $1.000\pm0.050$ & 29 \\
{\sii} & 16064.397 & -0.440 & 1.0 & 0.508 & 5.954 & 0.0 & 0.000 & 6.726 & $0.508\pm0.050$ & 4 \\
{\sii} & 16099.184 & -0.110 & 2.0 & 1.168 & 5.964 & 1.0 & 1.472 & 6.734 & $1.016\pm0.050$ & 19 \\
{\sii} & 16220.100 & -0.990 & 1.0 & 0.508 & 5.954 & 1.0 & 0.513 & 6.718 & $0.510\pm0.050$ & 2 \\
{\sii} & 16246.270 & -1.200 & 2.0 & 1.168 & 5.964 & 3.0 & 1.330 & 6.727 & $1.492\pm0.050$ & 23 \\
{\sii} & 16560.871 & -0.860 & 3.0 & 1.084 & 7.123 & 4.0 & 1.093 & 7.872 & $1.106\pm0.050$ & 1 \\
{\sii} & 16685.327 & -0.500 & 2.0 & 1.334 & 5.984 & 3.0 & 1.330 & 6.727 & $1.326\pm0.050$ & 38 \\
{\siii} & 16911.431 & 0.350 & 0.5 & 2.002 & 12.147 & 1.5 & 1.334 & 12.880 & $1.167\pm0.050$ & 18 \\
{\si} & 15408.000 & 0.520 & 2.0 & 1.167 & 8.699 & 3.0 & 1.084 & 9.504 & $1.001\pm0.050$ & 3 \\
{\si} & 15426.490 & 0.680 & 3.0 & 1.334 & 8.700 & 4.0 & 1.250 & 9.504 & $1.124\pm0.050$ & 10 \\
{\si} & 15482.712 & 0.000 & 2.0 & 1.501 & 8.046 & 1.0 & 2.002 & 8.846 & $1.250\pm0.050$ & 6 \\
{\ki} & 15167.211 & 0.640 & 2.5 & 1.143 & 2.670 & 3.5 & 1.200 & 3.487 & $1.271\pm0.050$ & 1 \\
{\cai} & 16155.175 & 0.362 & 1.0 & 1.496 & 4.532 & 2.0 & 1.166 & 5.300 & $1.001\pm0.050$ & 5 \\
{\cai} & 16161.778 & 0.492 & 1.0 & 1.005 & 4.554 & 2.0 & 1.001 & 5.321 & $0.999\pm0.050$ & 3 \\
{\cai} & 16201.500 & 0.638 & 2.0 & 1.501 & 4.535 & 3.0 & 1.334 & 5.300 & $1.167\pm0.050$ & 12 \\
{\caii} & 16114.659 & -0.796 & 1.5 & 0.800 & 9.983 & 2.5 & 0.857 & 10.753 & $0.900\pm0.050$ & 1 \\
{\caii} & 16244.130 & -0.790 & 1.5 & 1.334 & 10.107 & 2.5 & 1.200 & 10.870 & $1.099\pm0.050$ & 8 \\
{\caii} & 16565.589 & 0.366 & 0.5 & 0.666 & 9.235 & 1.5 & 0.800 & 9.983 & $0.834\pm0.050$ & 21 \\
{\caii} & 16654.425 & 0.626 & 1.5 & 1.334 & 9.240 & 2.5 & 1.200 & 9.984 & $1.099\pm0.050$ & 34 \\
{\caii} & 16667.278 & -0.334 & 1.5 & 1.334 & 9.240 & 1.5 & 0.800 & 9.983 & $1.067\pm0.050$ & 7 \\
{\tiii} & 15208.839 & -0.440 & 2.5 & 0.579 & 8.045 & 2.5 & 0.572 & 8.860 & $0.575\pm0.050$ & 1 \\
{\tiii} & 15734.183 & 0.438 & 6.5 & 1.231 & 8.132 & 5.5 & 1.273 & 8.920 & $1.116\pm0.050$ & 7 \\
{\tiii} & 15869.777 & 0.239 & 5.5 & 1.147 & 8.114 & 4.5 & 1.172 & 8.896 & $1.091\pm0.050$ & 1 \\
{\tiii} & 15878.172 & -1.925 & 2.5 & 1.200 & 3.124 & 3.5 & 1.147 & 3.904 & $1.081\pm0.050$ & 6 \\
{\tiii} & 15929.249 & 0.133 & 4.5 & 0.989 & 8.097 & 3.5 & 0.985 & 8.876 & $0.996\pm0.050$ & 1 \\
{\tiii} & 15936.446 & 0.076 & 3.5 & 0.684 & 8.082 & 2.5 & 0.572 & 8.860 & $0.824\pm0.050$ & 1 \\
{\tiii} & 16009.419 & -2.184 & 1.5 & 0.800 & 3.095 & 2.5 & 0.870 & 3.869 & $0.922\pm0.050$ & 3 \\
{\tiii} & 16190.997 & -0.230 & 3.5 & 1.144 & 7.866 & 2.5 & 1.200 & 8.632 & $1.074\pm0.050$ & 1 \\
{\tiii} & 16626.127 & -2.669 & 2.5 & 1.200 & 3.124 & 2.5 & 0.870 & 3.869 & $1.035\pm0.050$ & 2 \\
{\tiii} & 16727.485 & -0.048 & 4.5 & 1.111 & 8.409 & 3.5 & 1.143 & 9.150 & $1.055\pm0.050$ & 1 \\
{\cri} & 15684.347 & 0.270 & 2.0 & 2.002 & 4.697 & 3.0 & 1.680 & 5.487 & $1.358\pm0.050$ & 18 \\
{\cri} & 15978.569 & 0.400 & 2.0 & 1.237 & 5.978 & 3.0 & 1.279 & 6.754 & $1.321\pm0.050$ & 10 \\
{\crii} & 15158.083 & -0.509 & 4.5 & 1.275 & 10.768 & 4.5 & 1.436 & 11.586 & $1.355\pm0.050$ & 3 \\
{\crii} & 15165.729 & -0.561 & 3.5 & 1.442 & 10.893 & 3.5 & 1.631 & 11.711 & $1.536\pm0.050$ & 1 \\
{\crii} & 15180.104 & 0.807 & 2.5 & 1.326 & 13.600 & 3.5 & 1.202 & 14.417 & $1.047\pm0.050$ & 3 \\
{\crii} & 15189.927 & -0.706 & 2.5 & 1.590 & 10.850 & 1.5 & 1.948 & 11.667 & $1.322\pm0.050$ & 1 \\
{\crii} & 15204.755 & 0.934 & 3.5 & 1.404 & 13.615 & 4.5 & 1.298 & 14.430 & $1.113\pm0.050$ & 2 \\
{\crii} & 15252.203 & -0.726 & 3.5 & 1.546 & 10.860 & 2.5 & 1.786 & 11.672 & $1.246\pm0.050$ & 3 \\
{\crii} & 15256.391 & 1.115 & 5.5 & 1.455 & 13.657 & 6.5 & 1.385 & 14.470 & $1.192\pm0.050$ & 3 \\
{\crii} & 15293.817 & -0.455 & 3.5 & 1.590 & 10.754 & 3.5 & 1.948 & 11.565 & $1.769\pm0.050$ & 8 \\
{\crii} & 15369.672 & 0.399 & 6.5 & 1.385 & 10.804 & 5.5 & 1.455 & 11.610 & $1.192\pm0.050$ & 76 \\
{\crii} & 15373.877 & 0.671 & 3.5 & 1.675 & 13.650 & 4.5 & 1.488 & 14.457 & $1.161\pm0.050$ & 1 \\
{\crii} & 15375.106 & -0.741 & 4.5 & 1.458 & 10.904 & 3.5 & 1.631 & 11.711 & $1.155\pm0.050$ & 3 \\
{\crii} & 15389.066 & -0.857 & 1.5 & 1.756 & 10.845 & 0.5 & 3.310 & 11.651 & $1.367\pm0.050$ & 1 \\
{\crii} & 15409.033 & -0.495 & 2.5 & 0.861 & 10.744 & 2.5 & 1.317 & 11.548 & $1.089\pm0.050$ & 3 \\
{\crii} & 15470.129 & 0.289 & 5.5 & 1.344 & 10.784 & 4.5 & 1.436 & 11.586 & $1.137\pm0.050$ & 59 \\
{\crii} & 15519.564 & 0.282 & 2.5 & 1.672 & 13.655 & 1.5 & 1.799 & 14.454 & $1.577\pm0.050$ & 1 \\
{\crii} & 15520.215 & 0.825 & 4.5 & 1.546 & 13.686 & 5.5 & 1.446 & 14.485 & $1.221\pm0.050$ & 4 \\
{\crii} & 15552.511 & 0.164 & 4.5 & 1.275 & 10.768 & 3.5 & 1.399 & 11.565 & $1.058\pm0.050$ & 36 \\
{\crii} & 15611.316 & 0.019 & 3.5 & 1.146 & 10.754 & 2.5 & 1.317 & 11.548 & $0.932\pm0.050$ & 21 \\
{\crii} & 15641.791 & -0.147 & 2.5 & 0.861 & 10.744 & 1.5 & 1.069 & 11.536 & $0.705\pm0.050$ & 5 \\
{\crii} & 15926.280 & -0.251 & 4.5 & 1.533 & 10.872 & 3.5 & 1.587 & 11.650 & $1.438\pm0.050$ & 25 \\
{\crii} & 16054.895 & -0.234 & 4.5 & 1.458 & 10.904 & 4.5 & 1.554 & 11.676 & $1.506\pm0.050$ & 26 \\
{\crii} & 16159.919 & -0.382 & 3.5 & 1.546 & 10.860 & 2.5 & 1.620 & 11.627 & $1.454\pm0.050$ & 18 \\
{\crii} & 16192.181 & 0.018 & 5.5 & 1.454 & 10.911 & 4.5 & 1.554 & 11.676 & $1.229\pm0.050$ & 37 \\
{\crii} & 16372.266 & -0.397 & 3.5 & 1.186 & 10.893 & 3.5 & 1.620 & 11.650 & $1.403\pm0.050$ & 4 \\
{\crii} & 16617.092 & -0.625 & 4.5 & 1.458 & 10.904 & 3.5 & 1.587 & 11.650 & $1.232\pm0.050$ & 7 \\
{\crii} & 16619.219 & -0.608 & 2.5 & 1.389 & 10.881 & 2.5 & 1.620 & 11.627 & $1.505\pm0.050$ & 9 \\
{\mni} & 15157.808 & 0.036 & 2.5 & 2.000 & 6.127 & 3.5 & 1.716 & 6.945 & $1.361\pm0.050$ & 2 \\
{\mni} & 15221.919 & 0.507 & 3.5 & 2.000 & 4.889 & 3.5 & 1.938 & 5.703 & $1.969\pm0.050$ & 30 \\
{\mni} & 15266.671 & 0.379 & 3.5 & 2.000 & 4.889 & 2.5 & 2.288 & 5.701 & $1.640\pm0.050$ & 23 \\
{\mnii} & 15387.220 & -0.272 & 4.0 & 1.651 & 9.864 & 4.0 & 1.751 & 10.670 & $1.701\pm0.050$ & 5 \\
{\mnii} & 15412.667 & 0.237 & 5.0 & 1.601 & 9.865 & 4.0 & 1.751 & 10.670 & $1.301\pm0.050$ & 17 \\
{\mnii} & 15586.570 & -0.558 & 2.0 & 2.002 & 9.862 & 3.0 & 1.914 & 10.658 & $1.826\pm0.050$ & 3 \\
{\mnii} & 15600.576 & -0.140 & 3.0 & 1.752 & 9.863 & 3.0 & 1.914 & 10.658 & $1.833\pm0.050$ & 7 \\
{\mnii} & 15620.314 & -0.031 & 4.0 & 1.651 & 9.864 & 3.0 & 1.914 & 10.658 & $1.257\pm0.050$ & 8 \\
{\mnii} & 15737.053 & -0.303 & 1.0 & 3.004 & 9.862 & 2.0 & 2.293 & 10.650 & $1.938\pm0.050$ & 5 \\
{\mnii} & 15746.494 & -0.257 & 2.0 & 2.002 & 9.862 & 2.0 & 2.293 & 10.650 & $2.147\pm0.050$ & 4 \\
{\mnii} & 15760.789 & -0.479 & 3.0 & 1.752 & 9.863 & 2.0 & 2.293 & 10.650 & $1.211\pm0.050$ & 9 \\
{\fei} & 15211.686 & 0.231 & 3.0 & 1.755 & 5.385 & 2.0 & 1.983 & 6.200 & $1.527\pm0.050$ & 35 \\
{\fei} & 15223.777 & -0.086 & 3.0 & 1.508 & 5.587 & 2.0 & 1.468 & 6.401 & $1.548\pm0.050$ & 6 \\
{\fei} & 15243.880 & -0.020 & 3.0 & 1.492 & 6.419 & 2.0 & 1.397 & 7.232 & $1.587\pm0.050$ & 2 \\
{\fei} & 15249.140 & -0.277 & 3.0 & 1.508 & 5.587 & 3.0 & 1.285 & 6.400 & $1.397\pm0.050$ & 17 \\
{\fei} & 15298.742 & 0.622 & 5.0 & 1.585 & 5.308 & 5.0 & 1.595 & 6.119 & $1.590\pm0.050$ & 53 \\
{\fei} & 15339.578 & -0.007 & 2.0 & 2.009 & 5.410 & 1.0 & 2.963 & 6.218 & $1.532\pm0.050$ & 18 \\
{\fei} & 15505.559 & 0.007 & 5.0 & 1.360 & 6.286 & 6.0 & 1.263 & 7.086 & $1.020\pm0.050$ & 4 \\
{\fei} & 15592.522 & 0.323 & 4.0 & 1.487 & 6.367 & 4.0 & 1.493 & 7.162 & $1.490\pm0.050$ & 17 \\
{\fei} & 15595.753 & 0.356 & 6.0 & 1.490 & 6.242 & 7.0 & 1.357 & 7.037 & $0.958\pm0.050$ & 13 \\
{\fei} & 15608.489 & 0.005 & 6.0 & 1.490 & 6.242 & 6.0 & 1.433 & 7.036 & $1.462\pm0.050$ & 21 \\
{\fei} & 15617.894 & -0.785 & 3.0 & 1.294 & 6.350 & 4.0 & 1.219 & 7.144 & $1.107\pm0.050$ & 1 \\
{\fei} & 15625.923 & 0.468 & 4.0 & 1.502 & 5.539 & 4.0 & 1.486 & 6.333 & $1.494\pm0.050$ & 47 \\
{\fei} & 15636.221 & -0.019 & 4.0 & 1.655 & 5.352 & 4.0 & 1.653 & 6.144 & $1.654\pm0.050$ & 31 \\
{\fei} & 15657.151 & -0.019 & 5.0 & 1.508 & 6.246 & 5.0 & 1.285 & 7.038 & $1.397\pm0.050$ & 3 \\
{\fei} & 15666.296 & -0.858 & 5.0 & 1.421 & 5.828 & 4.0 & 1.392 & 6.619 & $1.479\pm0.050$ & 33 \\
{\fei} & 15681.805 & -0.200 & 5.0 & 1.510 & 6.246 & 5.0 & 1.357 & 7.037 & $1.433\pm0.050$ & 1 \\
{\fei} & 15690.729 & -0.197 & 5.0 & 1.510 & 6.246 & 5.0 & 1.448 & 7.036 & $1.479\pm0.050$ & 15 \\
{\fei} & 15696.145 & 0.133 & 5.0 & 1.510 & 6.246 & 6.0 & 1.433 & 7.036 & $1.241\pm0.050$ & 15 \\
{\fei} & 15727.882 & -0.485 & 2.0 & 1.503 & 5.621 & 3.0 & 1.645 & 6.409 & $1.787\pm0.050$ & 13 \\
{\fei} & 15773.732 & 0.604 & 4.0 & 1.502 & 5.539 & 5.0 & 1.402 & 6.325 & $1.202\pm0.050$ & 10 \\
{\fei} & 15778.381 & 0.343 & 4.0 & 1.574 & 6.299 & 5.0 & 1.412 & 7.085 & $1.088\pm0.050$ & 2 \\
{\fei} & 15802.879 & -0.099 & 4.0 & 1.514 & 6.252 & 5.0 & 1.448 & 7.036 & $1.316\pm0.050$ & 1 \\
{\fei} & 15908.671 & 0.029 & 4.0 & 1.338 & 6.365 & 5.0 & 1.348 & 7.144 & $1.368\pm0.050$ & 10 \\
{\fei} & 15910.390 & 0.289 & 2.0 & 1.503 & 5.621 & 3.0 & 1.285 & 6.400 & $1.067\pm0.050$ & 2 \\
{\fei} & 15915.651 & -0.405 & 4.0 & 1.331 & 5.874 & 3.0 & 1.210 & 6.653 & $1.512\pm0.050$ & 9 \\
{\fei} & 15924.995 & -0.239 & 4.0 & 1.585 & 6.258 & 5.0 & 1.357 & 7.037 & $0.901\pm0.050$ & 3 \\
{\fei} & 15985.094 & 0.440 & 6.0 & 1.351 & 6.264 & 7.0 & 1.292 & 7.040 & $1.115\pm0.050$ & 17 \\
{\fei} & 16042.203 & -0.248 & 6.0 & 1.351 & 6.264 & 6.0 & 1.514 & 7.037 & $1.433\pm0.050$ & 8 \\
{\fei} & 16045.040 & 0.247 & 4.0 & 1.331 & 5.874 & 4.0 & 1.409 & 6.647 & $1.370\pm0.050$ & 8 \\
{\fei} & 16047.093 & 0.028 & 3.0 & 1.615 & 6.265 & 3.0 & 1.508 & 7.038 & $1.562\pm0.050$ & 2 \\
{\fei} & 16106.810 & 0.402 & 4.0 & 1.331 & 5.874 & 5.0 & 1.275 & 6.644 & $1.163\pm0.050$ & 6 \\
{\fei} & 16169.447 & 0.467 & 6.0 & 1.415 & 6.319 & 7.0 & 1.362 & 7.086 & $1.203\pm0.050$ & 27 \\
{\fei} & 16208.682 & -0.333 & 4.0 & 1.238 & 6.321 & 4.0 & 1.283 & 7.086 & $1.260\pm0.050$ & 2 \\
{\fei} & 16212.175 & -0.041 & 4.0 & 1.238 & 6.321 & 5.0 & 1.274 & 7.085 & $1.346\pm0.050$ & 24 \\
{\fei} & 16217.970 & -0.278 & 3.0 & 1.687 & 6.275 & 3.0 & 1.644 & 7.039 & $1.665\pm0.050$ & 14 \\
{\fei} & 16230.054 & -0.317 & 3.0 & 1.244 & 6.380 & 4.0 & 1.219 & 7.144 & $1.182\pm0.050$ & 4 \\
{\fei} & 16236.084 & 0.464 & 4.0 & 1.355 & 6.380 & 5.0 & 1.348 & 7.144 & $1.334\pm0.050$ & 5 \\
{\fei} & 16289.221 & -0.961 & 3.0 & 1.318 & 6.398 & 4.0 & 1.441 & 7.159 & $1.626\pm0.050$ & 6 \\
{\fei} & 16320.781 & 0.555 & 7.0 & 1.429 & 6.280 & 8.0 & 1.376 & 7.040 & $1.190\pm0.050$ & 19 \\
{\fei} & 16386.732 & -0.139 & 7.0 & 1.429 & 6.280 & 7.0 & 1.402 & 7.037 & $1.415\pm0.050$ & 13 \\
{\fei} & 16491.173 & 0.708 & 5.0 & 1.421 & 5.828 & 6.0 & 1.334 & 6.580 & $1.117\pm0.050$ & 43 \\
{\fei} & 16521.738 & 0.051 & 5.0 & 1.360 & 6.286 & 6.0 & 1.306 & 7.037 & $1.171\pm0.050$ & 29 \\
{\fei} & 16528.983 & -0.516 & 5.0 & 1.384 & 6.336 & 6.0 & 1.263 & 7.086 & $0.960\pm0.050$ & 16 \\
{\fei} & 16536.502 & -0.552 & 5.0 & 1.360 & 6.286 & 5.0 & 1.448 & 7.036 & $1.404\pm0.050$ & 4 \\
{\fei} & 16556.519 & -0.269 & 2.0 & 0.967 & 6.411 & 3.0 & 1.029 & 7.159 & $1.091\pm0.050$ & 1 \\
{\fei} & 16617.302 & -0.211 & 3.0 & 1.318 & 6.398 & 4.0 & 1.219 & 7.144 & $1.071\pm0.050$ & 10 \\
{\fei} & 16650.424 & -0.059 & 2.0 & 0.991 & 5.956 & 2.0 & 1.137 & 6.700 & $1.064\pm0.050$ & 8 \\
{\fei} & 16665.933 & -0.059 & 3.0 & 1.499 & 6.342 & 4.0 & 1.283 & 7.086 & $0.959\pm0.050$ & 1 \\
{\srii} & 15214.400 & 0.033 & 3.5 & 1.143 & 7.562 & 2.5 & 1.200 & 8.377 & $1.072\pm0.050$ & 14 \\
{\srii} & 15310.900 & -0.018 & 0.0 & 0.857 & 7.562 & 1.5 & 0.800 & 8.372 & $0.775\pm0.050$ & 11 \\
{\ceiii} & 15720.131 & -3.080 & 4.0 & \nodata & 0.000 & 5.0 & \nodata & 0.789 & $1.390\pm0.106$ & 22 \\
{\ceiii} & 15961.157 & -1.120 & 4.0 & \nodata & 0.000 & 3.0 & \nodata & 0.777 & $0.852\pm0.056$ & 68 \\
{\ceiii} & 15964.928 & -1.660 & 2.0 & \nodata & 0.815 & 2.0 & \nodata & 1.591 & $0.970\pm0.153$ & 25 \\
{\ceiii} & 16133.170 & -0.920 & 6.0 & \nodata & 0.388 & 5.0 & \nodata & 1.156 & $1.123\pm0.048$ & 98 \\
{\ceiii} & 16292.642 & -2.430 & 2.0 & \nodata & 0.467 & 2.0 & \nodata & 1.227 & $0.929\pm0.053$ & 50 \\
???? & 15148.350 & \nodata & \nodata & \nodata & \nodata & \nodata & \nodata & \nodata & $0.996\pm0.099$ & 9 \\
???? & 15177.029 & \nodata & \nodata & \nodata & \nodata & \nodata & \nodata & \nodata & $1.076\pm0.075$ & 6 \\
???? & 15231.294 & \nodata & \nodata & \nodata & \nodata & \nodata & \nodata & \nodata & $1.187\pm0.067$ & 39 \\
???? & 15345.935 & \nodata & \nodata & \nodata & \nodata & \nodata & \nodata & \nodata & $1.308\pm0.052$ & 24 \\
???? & 15537.058 & \nodata & \nodata & \nodata & \nodata & \nodata & \nodata & \nodata & $1.310\pm0.070$ & 27 \\
???? & 15679.045 & \nodata & \nodata & \nodata & \nodata & \nodata & \nodata & \nodata & $0.922\pm0.092$ & 15 \\
???? & 15897.185 & \nodata & \nodata & \nodata & \nodata & \nodata & \nodata & \nodata & $1.210\pm0.152$ & 11 \\
???? & 16016.308 & \nodata & \nodata & \nodata & \nodata & \nodata & \nodata & \nodata & $1.084\pm0.082$ & 14 \\
???? & 16183.564 & \nodata & \nodata & \nodata & \nodata & \nodata & \nodata & \nodata & $1.049\pm0.059$ & 49 \\
???? & 16266.545 & \nodata & \nodata & \nodata & \nodata & \nodata & \nodata & \nodata & $1.046\pm0.069$ & 9 \\
???? & 16280.525 & \nodata & \nodata & \nodata & \nodata & \nodata & \nodata & \nodata & $1.016\pm0.065$ & 24 \\
???? & 16313.711 & \nodata & \nodata & \nodata & \nodata & \nodata & \nodata & \nodata & $0.802\pm0.086$ & 17 \\
???? & 16502.484 & \nodata & \nodata & \nodata & \nodata & \nodata & \nodata & \nodata & $1.195\pm0.050$ & 20 \\
???? & 16801.411 & \nodata & \nodata & \nodata & \nodata & \nodata & \nodata & \nodata & $1.055\pm0.035$ & 9 \\
???? & 16865.633 & \nodata & \nodata & \nodata & \nodata & \nodata & \nodata & \nodata & $0.903\pm0.051$ & 16 \\
\enddata
\end{deluxetable*}
\end{longrotatetable}

\clearpage

\begin{longrotatetable}
\begin{deluxetable*}{clrrcrlrrrr}
\tablecaption{Magnetic Field Modulus Estimates.}
\tabletypesize{\scriptsize}
\tablehead{
\colhead{2MASS ID} & \colhead{Other ID} & \colhead{$V$}   & \colhead{$H$}   & \colhead{$N_{\rm spectra}$} & \colhead{S/N} & \colhead{Spectral Type} & \colhead{{\bfield} (kG)} & \colhead{$N_{\rm lines}$} & \colhead{{\bfield} (kG)} & \colhead{$N_{\rm lines}$} \\
\colhead{}         & \colhead{}         & \colhead{(mag)} & \colhead{(mag)} & \colhead{}                  & \colhead{}    & \colhead{}           & \colhead{$H$-band}       & \colhead{$H$-band}        & \colhead{Optical}        & \colhead{Optical}        
}
\startdata
00033808+7018217 & HD 225114 & 8.10 & 8.19 & 12 & 726 & A0p SrCrSi\tablenotemark{4} & $7.46\pm0.54$ & 8 & $7.43\pm0.32$\tablenotemark{4} & 36 \\
00102704+7337035 & TYC 4306-1062-1 & 10.84 & 10.07 & 9 & 537 & \nodata & $4.17\pm0.28$ & 16 & \nodata & \nodata \\
00283062+6947472 & TYC 4299-696-1 & 10.53 & 9.45 & 12 & 682 & \nodata & $7.92\pm0.89$ & 4 & \nodata & \nodata \\
00284036+8418313 & TYC 4615-2915-1 & 10.96 & 10.54 & 14 & 332 & A2V\tablenotemark{2} & $5.10\pm0.22$ & 17 & \nodata & \nodata \\
00323366+5512530 & HD 2887 & 9.79 & 8.17 & 13 & 1793 & A2:p SrCr\tablenotemark{4} & $4.95\pm0.24$ & 25 & $4.97\pm0.22$\tablenotemark{4} & 27 \\
00331847+5716167 & TYC 3662-378-1 & 10.70 & 10.37 & 6 & 481 & \nodata & ($8.53\pm0.79$) & 5 & \nodata & \nodata \\
00564145+5739255 & TYC 3676-505-1 & 11.19 & 10.96 & 3 & 86 & \nodata & $18.49\pm0.76$ & 6 & \nodata & \nodata \\
00584870+6240562 & TYC 4021-632-1 & 11.22 & 10.45 & 6 & 411 & A2:p Eu\tablenotemark{4} & $4.25\pm0.29$ & 15 & $4.60\pm0.19$\tablenotemark{4} & 22 \\
01052242+5010296 & TYC 3271-1597-1 & 10.49 & 10.22 & 3 & 266 & \nodata & ($6.79\pm1.13$) & 3 & \nodata & \nodata \\
01184266+5844433 & TYC 3681-1528-1 & 10.49 & 9.99 & 4 & 355 & B8\tablenotemark{2} & $9.67\pm0.50$ & 9 & \nodata & \nodata \\
01225971$-$7335020 & HD 8700 & 9.83 & 9.42 & 3 & 327 & A0p SiCrFe\tablenotemark{1} & $8.69\pm0.33$ & 14 & \nodata & \nodata \\
02084669+6240520 & BD+62 352 & 10.16 & 9.57 & 6 & 543 & B8\tablenotemark{1} & ($7.34\pm1.16$) & 3 & \nodata & \nodata \\
02115402+3657583 & HD 13404 & 8.79 & 8.36 & 3 & 526 & A2p SrEu\tablenotemark{1} & $21.89\pm0.57$ & 17 & \nodata & \nodata \\
02210266+4256382 & HD 14437 & 7.25 & 7.29 & 3 & 1358 & B9p SrCrEu\tablenotemark{4} & $6.61\pm0.35$ & 17 & $7.58\pm0.05$\tablenotemark{5} & 1 \\
02252994+4759115 & HD 14873 & 9.12 & 8.48 & 7 & 951 & A5p SrCr\tablenotemark{4} & $5.63\pm0.35$ & 14 & $5.71\pm0.27$\tablenotemark{4} & 26 \\
02451791+6112352 & BD+60 562 & 10.11 & 9.39 & 1 & 171 & A0V\tablenotemark{2} & $9.88\pm0.35$ & 14 & \nodata & \nodata \\
02463240+0001145 & TYC 47-76-1 & 10.91 & 10.07 & 2 & 226 & \nodata & ($3.64\pm0.41$) & 10 & \nodata & \nodata \\
02511543+6447560 & BD+64 352 & 9.54 & 9.10 & 3 & 446 & B8p Si\tablenotemark{4} & $10.98\pm0.58$ & 6 & $10.56\pm0.43$\tablenotemark{4} & 43 \\
02563098+4534239 & TYC 3297-975-1 & 10.66 & 9.92 & 7 & 441 & A5:p EuSrCrSi\tablenotemark{4} & $24.61\pm0.72$ & 12 & $23.56\pm1.05$\tablenotemark{4} & 50 \\
02594527+5419449 & HD 18410 & 9.19 & 8.56 & 3 & 574 & A2p SiCrEu\tablenotemark{1} & ($5.01\pm0.41$) & 11 & \nodata & \nodata \\
03001958+4355254 & TYC 2859-794-1 & 11.36 & 10.51 & 7 & 332 & \nodata & $5.33\pm0.21$ & 17 & \nodata & \nodata \\
03010340+5816543 & HD 237049 & 9.34 & 8.58 & 7 & 871 & A2\tablenotemark{2} & ($5.07\pm0.56$) & 7 & \nodata & \nodata \\
03080905+4355586 & TYC 2859-1418-1 & 11.19 & 10.27 & 7 & 340 & \nodata & $5.91\pm0.18$ & 37 & \nodata & \nodata \\
03112102+6310394 & TYC 4053-630-1 & 10.17 & 9.89 & 3 & 301 & B9p SiCr\tablenotemark{4} & $14.52\pm0.63$ & 7 & $13.10\pm0.58$\tablenotemark{4} & 50 \\
03302967+4715089 & TYC 3316-77-1 & 11.37 & 10.97 & 14 & 362 & A2\tablenotemark{2} & $7.81\pm0.43$ & 9 & \nodata & \nodata \\
03414761+2335305 & TYC 1799-1426-1 & 11.41 & 9.87 & 3 & 252 & A6p EuSrSiCr\tablenotemark{4} & $9.61\pm0.17$ & 54 & $9.66\pm0.39$\tablenotemark{4} & 50 \\
04011662+4655136 & HD 25092 & 9.43 & 8.86 & 3 & 258 & A3p EuSrCrSi\tablenotemark{4} & $6.09\pm0.40$ & 11 & $6.47\pm0.29$\tablenotemark{4} & 31 \\
04013793+5736159 & BD+57 764 & 10.00 & 9.56 & 3 & 217 & B9p Si\tablenotemark{2} & ($5.60\pm0.98$) & 4 & \nodata & \nodata \\
04021373+3709570 & HD 279277 & 10.70 & 9.88 & 7 & 541 & B8V\tablenotemark{2} & $8.10\pm0.62$ & 11 & \nodata & \nodata \\
04064234+4546408 & HD 25706 & 10.19 & 9.62 & 3 & 329 & A0p Si\tablenotemark{1} & $9.33\pm0.50$ & 9 & \nodata & \nodata \\
04093719+5302064 & TYC 3718-1499-1 & 11.13 & 10.56 & 12 & 546 & \nodata & $5.33\pm0.30$ & 12 & \nodata & \nodata \\
04140370+4941107 & TYC 3336-982-1 & 11.90 & 10.91 & 6 & 237 & \nodata & ($5.46\pm0.76$) & 5 & \nodata & \nodata \\
04203778+2853313 & HD 27404 & 8.78 & 7.28 & 6 & 1225 & A0p Si\tablenotemark{1} & ($5.29\pm0.46$) & 13 & \nodata & \nodata \\
04294199+3041025 & HD 282151 & 10.65 & 9.89 & 6 & 500 & B9\tablenotemark{3} & $5.41\pm0.41$ & 14 & \nodata & \nodata \\
04434133+6356292 & TYC 4086-1878-1 & 11.15 & 10.50 & 9 & 405 & \nodata & ($5.51\pm0.47$) & 12 & \nodata & \nodata \\
04580135+3007324 & HD 31552 & 8.95 & 7.46 & 1 & 527 & A0\tablenotemark{3} & $4.16\pm0.44$ & 10 & \nodata & \nodata \\
05013296+6301162 & HD 31629 & 9.29 & 8.79 & 9 & 696 & A0\tablenotemark{3} & $7.92\pm0.65$ & 8 & \nodata & \nodata \\
05060835+3355072 & HD 32633 & 7.05 & 7.03 & 3 & 1070 & B9p SiCr\tablenotemark{1} & $10.25\pm1.00$ & 4 & \nodata & \nodata \\
05145454+1605118 & HD 241957 & 9.88 & 8.95 & 3 & 275 & A3\tablenotemark{2} & ($5.00\pm0.46$) & 13 & \nodata & \nodata \\
05151044+3300479 & HD 241843 & 10.33 & 9.70 & 3 & 385 & A0/2\tablenotemark{2} & $3.90\pm0.33$ & 11 & \nodata & \nodata \\
05225493+1346388 & HD 243096 & 11.17 & 10.22 & 7 & 420 & A5\tablenotemark{2} & ($5.80\pm0.65$) & 6 & \nodata & \nodata \\
05253809+3041142 & HD 35379 & 8.90 & 8.83 & 3 & 525 & B9p SiSr\tablenotemark{1} & $5.71\pm0.91$ & 4 & \nodata & \nodata \\
05254840+6216036 & TYC 4085-1776-1 & 11.20 & 10.70 & 3 & 98 & \nodata & $5.18\pm0.19$ & 30 & \nodata & \nodata \\
05331986+4013456 & TYC 2914-1134-1 & 10.48 & 9.96 & 3 & 158 & \nodata & $4.63\pm0.41$ & 9 & \nodata & \nodata \\
05332350+2606572 & TYC 1852-651-1 & 11.69 & 10.78 & 14 & 424 & \nodata & $5.10\pm0.53$ & 6 & \nodata & \nodata \\
05334537+1238120 & HD 36644 & 9.57 & 9.10 & 4 & 467 & A0\tablenotemark{3} & $18.48\pm0.31$ & 25 & \nodata & \nodata \\
05363639+3055204 & TYC 2404-880-1 & 11.04 & 10.48 & 6 & 338 & B8-A2\tablenotemark{2} & $6.27\pm0.27$ & 15 & \nodata & \nodata \\
05451425+3654510 & TYC 2417-4-1 & 11.36 & 10.92 & 3 & 94 & \nodata & $10.62\pm0.42$ & 19 & \nodata & \nodata \\
05472522+1238532 & HD 38586 & 9.33 & 8.76 & 9 & 1059 & B9\tablenotemark{3} & $6.73\pm0.67$ & 5 & \nodata & \nodata \\
05481976+3335169 & HD 247628 & 10.50 & 10.07 & 3 & 122 & A5p SiSr\tablenotemark{1} & $8.79\pm0.22$ & 33 & \nodata & \nodata \\
05482124+0305160 & TYC 120-185-1 & 11.51 & 10.35 & 3 & 215 & \nodata & $4.63\pm0.33$ & 12 & \nodata & \nodata \\
05482549$-$0045346 & HD 38823 & 8.88 & 6.95 & 6 & 1885 & A5p SrEuCr\tablenotemark{1} & $5.13\pm0.35$ & 12 & \nodata & \nodata \\
05510825+5316108 & TYC 3750-928-1 & 11.53 & 10.96 & 7 & 275 & \nodata & $5.37\pm0.20$ & 26 & \nodata & \nodata \\
05555073$-$7750438 & HD 41613 & 9.81 & 8.89 & 2 & 295 & A3p EuCr\tablenotemark{1} & $5.41\pm0.37$ & 17 & \nodata & \nodata \\
06010117+3214538 & TYC 2423-267-1 & 10.78 & 10.30 & 14 & 512 & \nodata & $9.87\pm0.21$ & 37 & \nodata & \nodata \\
06055681+1917131 & TYC 1321-1213-1 & 10.58 & 10.33 & 9 & 361 & \nodata & $5.76\pm0.26$ & 15 & \nodata & \nodata \\
06062179+2124358 & HD 251556 & 10.26 & 10.24 & 6 & 396 & B9V\tablenotemark{3} & $6.83\pm0.53$ & 5 & \nodata & \nodata \\
06070875+1053283 & TYC 721-523-1 & 10.92 & 10.65 & 7 & 230 & \nodata & $8.54\pm1.17$ & 3 & \nodata & \nodata \\
06091241+2026527 & HD 252382 & 10.63 & 10.45 & 9 & 391 & B6III\tablenotemark{3} & $5.60\pm0.57$ & 7 & \nodata & \nodata \\
06130161+7413402 & TYC 4357-491-1 & 10.68 & 10.20 & 4 & 274 & \nodata & $5.31\pm0.30$ & 19 & \nodata & \nodata \\
06163561$-$1829384 & TYC 5938-2491-1 & 10.80 & 10.05 & 2 & 304 & \nodata & $10.15\pm0.35$ & 11 & \nodata & \nodata \\
06171773$-$0214319 & HD 291513 & 10.62 & 10.39 & 12 & 402 & A2\tablenotemark{2} & $5.25\pm0.48$ & 6 & \nodata & \nodata \\
06285110+5412297 & TYC 3765-681-1 & 10.81 & 10.55 & 3 & 189 & \nodata & ($9.27\pm1.28$) & 3 & \nodata & \nodata \\
06340185+3331392 & HD 46297 & 8.97 & 8.39 & 9 & 1185 & A2\tablenotemark{3} & $6.06\pm0.23$ & 36 & \nodata & \nodata \\
06352208+3351329 & TYC 2430-1205-1 & 10.86 & 10.41 & 9 & 450 & \nodata & $5.63\pm0.25$ & 27 & \nodata & \nodata \\
06365367+0118455 & HD 47074 & 9.43 & 9.33 & 12 & 881 & A2p SrCrEu\tablenotemark{1} & $11.85\pm0.15$ & 61 & \nodata & \nodata \\
06411268+2405034 & HD 47774 & 9.07 & 8.61 & 6 & 900 & A0\tablenotemark{3} & $8.78\pm1.36$ & 2 & \nodata & \nodata \\
06444568+1135345 & HD 263064 & 10.34 & 10.10 & 6 & 414 & B8V\tablenotemark{2} & ($5.39\pm0.51$) & 6 & \nodata & \nodata \\
06515922$-$0138404 & HD 50169 & 9.35 & 8.95 & 8 & 698 & A3p SrEuCr\tablenotemark{1} & $4.40\pm0.26$ & 30 & $5.03\pm0.03$\tablenotemark{5} & 1 \\
06545817+0408276 & HD 266311 & 9.74 & 9.73 & 15 & 345 & A2p SrCrEu\tablenotemark{1} & $4.43\pm0.25$ & 24 & \nodata & \nodata \\
06550983$-$3003440 & HD 51203 & 10.39 & 9.89 & 4 & 413 & A2p SrEuCr\tablenotemark{1} & $7.90\pm0.51$ & 6 & \nodata & \nodata \\
07051562+5542284 & HD 52628 & 8.66 & 7.99 & 3 & 726 & A2p CrEu\tablenotemark{1} & $6.68\pm0.18$ & 53 & \nodata & \nodata \\
07123042$-$2103537 & HD 55540 & 9.52 & 9.37 & 3 & 398 & A0p EuCr\tablenotemark{1} & $15.03\pm0.89$ & 7 & $12.73\pm0.30$\tablenotemark{6} & 1 \\
07141263$-$0341299 & HD 296704 & 10.22 & 10.27 & 3 & 235 & B8\tablenotemark{2} & ($4.69\pm0.50$) & 8 & \nodata & \nodata \\
07163050+1519123 & TYC 1346-358-1 & 11.21 & 10.86 & 3 & 188 & A3p EuSrCrSi\tablenotemark{4} & $17.20\pm0.39$ & 13 & $17.11\pm0.60$\tablenotemark{4} & 50 \\
07175283+1347078 & HD 56514 & 9.63 & 9.05 & 3 & 522 & A5:III:\tablenotemark{2} & $6.60\pm0.45$ & 8 & \nodata & \nodata \\
07194142+1448162 & TYC 775-617-1 & 11.03 & 10.60 & 3 & 122 & \nodata & $7.05\pm0.74$ & 5 & \nodata & \nodata \\
07262241$-$6026516 & TYC 8910-1832-1 & 11.03 & 10.10 & 1 & 105 & \nodata & $6.71\pm0.26$ & 19 & \nodata & \nodata \\
07320894$-$5929216 & TYC 8563-764-1 & 11.11 & 10.53 & 1 & 79 & \nodata & $6.61\pm0.70$ & 4 & \nodata & \nodata \\
07320905+3922431 & TYC 2962-181-1 & 11.36 & 10.77 & 5 & 196 & A5p EuSrCr\tablenotemark{4} & $9.26\pm0.31$ & 18 & $9.04\pm0.40$\tablenotemark{4} & 50 \\
07390490+4711571 & BD+47 1470 & 9.66 & 9.33 & 2 & 413 & A0\tablenotemark{3} & ($5.25\pm0.37$) & 10 & \nodata & \nodata \\
07470594+5338193 & BD+53 1183 & 9.98 & 9.70 & 3 & 297 & Ap CrSrEu\tablenotemark{1} & $5.34\pm0.59$ & 9 & \nodata & \nodata \\
07524109+4350146 & HD 63813 & 9.71 & 9.19 & 2 & 342 & A1-F0 Sr\tablenotemark{1} & ($5.40\pm0.34$) & 15 & \nodata & \nodata \\
07525522$-$5216518 & TYC 8147-1121-1 & 11.69 & 10.95 & 1 & 110 & \nodata & $13.08\pm0.60$ & 14 & \nodata & \nodata \\
07583427+2035316 & HD 65240 & 9.13 & 8.43 & 1 & 442 & A0\tablenotemark{3} & ($4.97\pm0.40$) & 11 & \nodata & \nodata \\
08053890+5121205 & TYC 3414-366-1 & 10.30 & 9.70 & 2 & 284 & \nodata & $5.87\pm0.36$ & 13 & \nodata & \nodata \\
08124763$-$1219281 & HD 68619 & 9.96 & 9.46 & 3 & 392 & A8/9V\tablenotemark{2} & $4.36\pm0.31$ & 14 & \nodata & \nodata \\
08171198+4018115 & TYC 2977-1620-1 & 11.11 & 10.61 & 3 & 263 & \nodata & $6.00\pm0.30$ & 22 & \nodata & \nodata \\
08273117$-$4931353 & HD 71860 & 9.88 & 9.18 & 1 & 294 & A0p SiCr\tablenotemark{1} & $8.87\pm0.29$ & 20 & \nodata & \nodata \\
11193525+0955569 & HD 98437 & 9.39 & 8.26 & 1 & 298 & A3\tablenotemark{3} & ($3.58\pm0.33$) & 31 & \nodata & \nodata \\
12020633+1520417 & HD 104505 & 9.74 & 9.51 & 3 & 272 & A5p EuSr\tablenotemark{4} & $7.59\pm0.22$ & 39 & $8.02\pm0.28$\tablenotemark{4} & 50 \\
12231298$-$6316319 & TYC 8979-339-1 & 10.41 & 9.69 & 16 & 879 & \nodata & $7.96\pm0.17$ & 43 & \nodata & \nodata \\
13402136+5712275 & HD 119213 & 6.30 & 6.21 & 3 & 1293 & A3p SrCr\tablenotemark{1} & $6.75\pm0.98$ & 4 & \nodata & \nodata \\
15404512$-$4004586 & HD 139631 & 8.76 & 8.15 & 1 & 541 & A0p EuCrSr\tablenotemark{1} & $9.06\pm0.45$ & 11 & \nodata & \nodata \\
16015886$-$3732036 & HD 143473 & 8.86 & 6.98 & 1 & 865 & B9p Si\tablenotemark{1} & ($9.56\pm1.19$) & 4 & \nodata & \nodata \\
16241403$-$4044107 & HD 325559 & 10.33 & 8.51 & 9 & 1138 & A0\tablenotemark{2} & ($8.17\pm1.15$) & 3 & \nodata & \nodata \\
16505838$-$6329330 & CD-63 1232 & 9.73 & 9.78 & 5 & 395 & \nodata & ($8.33\pm1.09$) & 3 & \nodata & \nodata \\
16595976+1002298 & TYC 980-1372-1 & 10.91 & 10.35 & 4 & 279 & \nodata & ($6.46\pm0.53$) & 8 & \nodata & \nodata \\
17080901$-$6045032 & HD 154253 & 9.02 & 8.56 & 1 & 299 & A0p SrCrEu\tablenotemark{1} & ($5.67\pm0.58$) & 5 & \nodata & \nodata \\
17293722$-$3030290 & HD 317652 & 11.78 & 10.91 & 1 & 54 & A\tablenotemark{3} & $7.19\pm0.48$ & 6 & \nodata & \nodata \\
17355458$-$3350564 & HD 159379 & 8.56 & 8.14 & 2 & 665 & B9p Si\tablenotemark{1} & ($5.90\pm1.12$) & 3 & \nodata & \nodata \\
17380465$-$2146101 & UCAC4 342-102489 & 11.89 & 10.88 & 1 & 133 & \nodata & $4.62\pm0.32$ & 11 & \nodata & \nodata \\
17534913$-$3332308 & HD 318569 & 10.42 & 9.88 & 3 & 194 & A2\tablenotemark{3} & $7.37\pm0.63$ & 6 & \nodata & \nodata \\
17592320$-$2624123 & HD 163850 & 9.84 & 9.28 & 1 & 171 & B8\tablenotemark{2} & ($7.63\pm1.17$) & 3 & \nodata & \nodata \\
17594524+0302056 & TYC 421-935-1 & 11.01 & 10.50 & 3 & 276 & \nodata & $17.13\pm0.33$ & 17 & \nodata & \nodata \\
18010675$-$3137300 & HD 318820 & 10.24 & 9.97 & 4 & 377 & A0\tablenotemark{2} & $5.05\pm0.62$ & 6 & \nodata & \nodata \\
18061507+0237308 & HD 165525 & 10.27 & 10.04 & 3 & 132 & A2\tablenotemark{2} & $5.89\pm0.52$ & 6 & \nodata & \nodata \\
18133046$-$2815189 & HD 166808 & 9.14 & 8.91 & 3 & 381 & A2p SrCr\tablenotemark{1} & $5.02\pm0.35$ & 12 & \nodata & \nodata \\
18151324+0121014 & UCAC4 457-075953 & 12.11 & 10.30 & 16 & 589 & \nodata & $4.63\pm0.26$ & 19 & \nodata & \nodata \\
18171434+0013052 & \nodata & 12.34 & 10.82 & 16 & 371 & \nodata & $10.53\pm0.30$ & 15 & \nodata & \nodata \\
18261274$-$0646548 & BD-06 4760 & 10.24 & 8.33 & 3 & 300 & A3V\tablenotemark{2} & ($4.64\pm0.25$) & 18 & \nodata & \nodata \\
18300828$-$0235273 & HD 170565 & 9.23 & 8.49 & 6 & 551 & A5p SrEuCrSi\tablenotemark{4} & $6.56\pm0.34$ & 21 & $6.93\pm0.25$\tablenotemark{4} & 29 \\
18305040$-$2402152 & TYC 3895-1263-1 & 10.69 & 10.21 & 1 & 76 & \nodata & ($5.59\pm1.00$) & 4 & \nodata & \nodata \\
18325156+1718124 & BD+17 3622 & 8.85 & 8.69 & 3 & 522 & A2p SrEuCr\tablenotemark{1} & $4.06\pm0.24$ & 12 & \nodata & \nodata \\
18521845+1437019 & BD+14 3679 & 9.64 & 8.89 & 3 & 469 & A0\tablenotemark{3} & $6.03\pm0.40$ & 11 & \nodata & \nodata \\
19013248+4151592 & HD 177128 & 9.19 & 9.05 & 3 & 409 & A1Vp SiCrCrEu\tablenotemark{2} & $5.00\pm0.25$ & 16 & \nodata & \nodata \\
19120493+4817166 & KIC 10852970 & 10.57 & 10.27 & 3 & 204 & A2Vp\tablenotemark{3} & $6.22\pm0.36$ & 9 & \nodata & \nodata \\
19130947+1201217 & HD 179711 & 8.36 & 8.31 & 4 & 677 & A0p Si\tablenotemark{1} & ($6.81\pm0.53$) & 11 & \nodata & \nodata \\
19152589$-$1142537 & HD 180058 & 9.80 & 8.86 & 1 & 377 & A3p Sr\tablenotemark{1} & ($6.80\pm0.61$) & 5 & \nodata & \nodata \\
19183796+3947264 & KIC 4647715 & 10.99 & 10.69 & 5 & 255 & A1Vp\tablenotemark{3} & $6.33\pm0.47$ & 8 & \nodata & \nodata \\
19191030$-$0654130 & TYC 5142-2803-1 & 11.45 & 10.69 & 13 & 239 & \nodata & ($5.90\pm0.38$) & 9 & \nodata & \nodata \\
19192992+1504587 & HD 231151 & 10.32 & 9.63 & 5 & 534 & B9\tablenotemark{2} & $19.40\pm0.44$ & 7 & \nodata & \nodata \\
19195551+3652567 & KIC 1157401 & 9.88 & 9.54 & 3 & 387 & \nodata & ($8.27\pm0.60$) & 5 & \nodata & \nodata \\
19221039+5048401 & KIC 12207099 & 10.20 & 10.31 & 2 & 219 & A0\tablenotemark{2} & ($6.08\pm0.43$) & 12 & \nodata & \nodata \\
19224960$-$1335307 & BD-13 5332 & 10.66 & 10.40 & 6 & 320 & Ap Si\tablenotemark{1} & ($5.91\pm0.52$) & 7 & \nodata & \nodata \\
19265158$-$0038427 & TYC 5131-1737-1 & 11.00 & 10.39 & 7 & 340 & A0\tablenotemark{2} & $4.36\pm0.23$ & 28 & \nodata & \nodata \\
19292391$-$0857083 & TYC 5722-1600-1 & 11.98 & 10.97 & 3 & 174 & \nodata & ($5.54\pm0.86$) & 4 & \nodata & \nodata \\
19304169+1621445 & HD 231627 & 10.82 & 10.05 & 3 & 258 & A\tablenotemark{2} & ($6.57\pm1.33$) & 2 & \nodata & \nodata \\
19305065+0102421 & HD 183735 & 9.96 & 9.01 & 7 & 748 & A2p SrCrEu\tablenotemark{1} & $5.75\pm0.27$ & 21 & \nodata & \nodata \\
19443776+1432091 & HD 353513 & 10.80 & 10.52 & 3 & 200 & A1\tablenotemark{3} & ($5.64\pm0.65$) & 7 & \nodata & \nodata \\
19462921+4737500 & KIC 10483436 & 11.39 & 10.88 & 7 & 265 & Ap\tablenotemark{3} & ($3.74\pm0.35$) & 7 & \nodata & \nodata \\
19510964+1827101 & HD 350689 & 11.40 & 10.90 & 6 & 172 & B9\tablenotemark{3} & $5.15\pm0.48$ & 13 & \nodata & \nodata \\
19525699+1748543 & HD 188103 & 8.02 & 8.13 & 19 & 1743 & A0p SiSrEu\tablenotemark{4} & $10.10\pm0.62$ & 11 & $10.74\pm0.43$\tablenotemark{4} & 50 \\
20020903+0753287 & TYC 1071-2065-1 & 11.20 & 10.91 & 7 & 274 & \nodata & ($4.60\pm0.21$) & 17 & \nodata & \nodata \\
20191831+2354023 & HD 346258 & 9.94 & 9.65 & 3 & 361 & A2\tablenotemark{3} & $4.76\pm0.21$ & 27 & \nodata & \nodata \\
21054288+4035346 & BD+40 4409 & 9.94 & 9.52 & 3 & 356 & A4p EuSrCrSi\tablenotemark{4} & $9.42\pm0.21$ & 44 & $9.71\pm0.38$\tablenotemark{4} & 50 \\
21063689+4837401 & HD 201250 & 8.84 & 8.79 & 25 & 1757 & B8p Si\tablenotemark{4} & $7.81\pm0.69$ & 6 & $7.81\pm0.36$\tablenotemark{4} & 36 \\
21090244+4844449 & HD 201612 & 8.57 & 8.72 & 25 & 1838 & B9p SiCrEu:\tablenotemark{4} & $6.74\pm0.49$ & 11 & $7.43\pm0.34$\tablenotemark{4} & 45 \\
21135681+4702336 & TYC 3593-1718-1 & 11.43 & 10.90 & 25 & 514 & \nodata & $4.76\pm0.26$ & 24 & \nodata & \nodata \\
21220944+5640490 & HD 203763 & 9.42 & 8.96 & 3 & 639 & A4:p SrCrEu\tablenotemark{4} & $4.72\pm0.33$ & 17 & $5.04\pm0.25$\tablenotemark{4} & 17 \\
21282592+4647232 & TYC 3590-1739-1 & 10.78 & 10.30 & 16 & 541 & \nodata & ($6.29\pm0.80$) & 6 & \nodata & \nodata \\
21402960+4326586 & TYC 3196-875-1 & 10.77 & 10.59 & 12 & 505 & \nodata & $6.81\pm0.40$ & 10 & \nodata & \nodata \\
21432172+4151293 & TYC 3192-1815-1 & 11.10 & 10.66 & 12 & 485 & \nodata & ($4.88\pm0.48$) & 7 & \nodata & \nodata \\
21460991+5247363 & TYC 3967-1331-1 & 10.85 & 10.52 & 3 & 301 & B:\tablenotemark{2} & ($4.65\pm1.37$) & 3 & \nodata & \nodata \\
21480795+5802409 & BD+57 2400 & 10.32 & 9.63 & 3 & 388 & A5p EuSrCrSi\tablenotemark{4} & $14.64\pm0.74$ & 9 & $15.12\pm0.61$\tablenotemark{4} & 50 \\
21535503+5417550 & HD 208325 & 9.13 & 8.52 & 14 & 1234 & A4p Sr\tablenotemark{4} & $4.58\pm0.42$ & 12 & $5.12\pm0.21$\tablenotemark{4} & 23 \\
22054630+4913580 & HD 209931 & 9.40 & 9.07 & 3 & 445 & A0p SiCrSr\tablenotemark{4} & $5.35\pm0.63$ & 6 & $5.91\pm0.25$\tablenotemark{4} & 20 \\
22094501+6154365 & HD 210626 & 9.21 & 8.69 & 3 & 499 & A0p CrSr\tablenotemark{4} & $6.31\pm0.50$ & 9 & $6.29\pm0.32$\tablenotemark{4} & 36 \\
22102021+6618150 & HD 210759 & 8.46 & 7.83 & 5 & 753 & A2p CrSrEu\tablenotemark{4} & $4.72\pm0.23$ & 17 & $5.10\pm0.22$\tablenotemark{4} & 24 \\
22430180+5141123 & HD 235936 & 9.61 & 9.47 & 3 & 310 & B9p CrSiSr\tablenotemark{4} & $5.09\pm1.17$ & 2 & $4.90\pm0.22$\tablenotemark{4} & 15 \\
22480641+5906428 & HD 216001 & 8.60 & 8.27 & 6 & 955 & B8p SrSiCr\tablenotemark{4} & $6.99\pm0.51$ & 9 & $7.75\pm0.35$\tablenotemark{4} & 31 \\
23102121+4717017 & BD+46 3957 & 10.72 & 10.53 & 3 & 236 & A3:p SrEuCrSi\tablenotemark{4} & $18.36\pm0.70$ & 12 & $17.91\pm0.77$\tablenotemark{4} & 50 \\
23222946+6009121 & \nodata & 10.44 & 10.22 & 3 & 326 & A0p CrEuSrSi\tablenotemark{4} & $6.18\pm0.52$ & 5 & $6.72\pm0.30$\tablenotemark{4} & 50 \\
23223523+6201042 & BD+61 2436 & 9.75 & 9.47 & 3 & 382 & B9p CrSrSi\tablenotemark{4} & $7.39\pm0.58$ & 7 & $7.53\pm0.28$\tablenotemark{4} & 45 \\
23270758+4933290 & TYC 3645-984-1 & 10.54 & 10.30 & 6 & 445 & A0p SrCr\tablenotemark{4} & $6.06\pm0.49$ & 8 & $6.22\pm0.27$\tablenotemark{4} & 36 \\
23540533+6616511 & BD+65 1962 & 10.69 & 9.94 & 3 & 319 & \nodata & $6.52\pm0.73$ & 4 & \nodata & \nodata \\
23554319+5647321 & BD+55 3046 & 9.44 & 9.15 & 4 & 559 & B9p SiSr\tablenotemark{1} & $5.01\pm0.36$ & 13 & \nodata & \nodata \\
\enddata
\tablenotetext{1}{\citet{renson09}}
\tablenotetext{2}{\citet{skiff14}}
\tablenotetext{3}{SIMBAD}
\tablenotetext{4}{Own data (ARC 3.5m/ARCES)}
\tablenotetext{5}{\citet{mathys17}}
\tablenotetext{6}{\citet{freyhammer08}}
\end{deluxetable*}
\end{longrotatetable}

\end{document}